\documentclass[aoas,preprint]{imsart}
\usepackage{amsmath}
\usepackage{graphicx}
\usepackage{enumerate}
\usepackage{natbib}
\usepackage{url} % not crucial - just used below for the URL 

\usepackage{graphicx}
\usepackage{wrapfig}
\usepackage{lscape}
\usepackage{rotating}
\usepackage{epstopdf}
\usepackage{etoolbox}

\newcommand{\phik}{\phi_k}

\newcommand{\cIRk}{\mbox{cloglog}(\mbox{IR}_k)}
\newcommand{\cIFRk}{\mbox{cloglog}(\mbox{IFR}_k)}

\usepackage{xcolor}

\begin{document}

\begin{frontmatter}
\title{Bayesian adjustment for preferential testing in estimating infection fatality rates, as motivated by the COVID-19 pandemic}
%\title{A sample article title with some additional note\thanksref{t1}}
\runtitle{Bayesian adjustment for preferential testing in estimating the COVID-19 IFR}
%\thankstext{T1}{A sample additional note to the title.}

\begin{aug}
%%%%%%%%%%%%%%%%%%%%%%%%%%%%%%%%%%%%%%%%%%%%%%
%%Only one address is permitted per author. %%
%%Only division, organization and e-mail is %%
%%included in the address.                  %%
%%Additional information can be included in %%
%%the Acknowledgments section if necessary. %%
%%%%%%%%%%%%%%%%%%%%%%%%%%%%%%%%%%%%%%%%%%%%%%
\author{Harlan Campbell$^{1,}$ \thanks{1. Department of Statistics, University of British Columbia, BC, Canada;  2. 
Department of Environmental Science,
Policy, and Management, University of
California, Berkeley, CA, USA; 3. Heidelberg Institute for Global Health, Heidelberg University Hospital, Heidelberg, Germany; 4. Julius Center for Health Sciences and Primary Care, University Medical Center Utrecht, Utrecht University, Utrecht, the Netherlands; 5. Cochrane Netherlands, Julius Center for Health Sciences and Primary Care, University Medical Center Utrecht, Utrecht University, Utrecht the Netherlands; 6. Department of Epidemiology, Colorado School of Public Health, CO, USA; This work was supported by the European Union's Horizon 2020 research and innovation programme under ReCoDID grant agreement No 825746 and by the Canadian Institutes of Health Research, Institute of Genetics (CIHR-IG) under Grant Agreement N$^{o}$ 01886-000.  We also wish to thank Joe Watson for his input early on and expertise on preferential sampling,}
Perry {de Valpine}$^{2}$, 
Lauren Maxwell$^{3}$, \\
Valentijn M.T. {de Jong}$^{4}$, 
Thomas P.A. Debray$^{4,5}$, \\
Thomas Jaenisch$^{3,6}$, 
Paul Gustafson$^{1}$}
%%%%%%%%%%%%%%%%%%%%%%%%%%%%%%%%%%%%%%%%%%%%%%
%% Addresses                                %%
%%%%%%%%%%%%%%%%%%%%%%%%%%%%%%%%%%%%%%%%%%%%%%

\end{aug}

\begin{abstract}
A key challenge in estimating the infection fatality rate (IFR) \textcolor{black}{-and its relation with various factors of interest-} is determining the total number of cases.  The total number of cases is not known because not everyone is tested, but also, more importantly, because tested individuals are not representative of the population at large.  We refer to the phenomenon whereby infected individuals are more likely to be tested than non-infected individuals,  as ``preferential testing.''  An open question is whether or not it is possible to reliably estimate the IFR without any specific knowledge about the degree to which the data are biased by preferential testing. In this paper we take a partial identifiability approach, formulating clearly where deliberate prior assumptions can be made and presenting a Bayesian model which pools information from different samples.  When the model is fit to European data obtained from seroprevalence studies and national official COVID-19 statistics, we estimate the overall COVID-19 IFR for Europe to be 0.53\%,  95\% C.I.    =  [0.39\%,  0.69\%].
\end{abstract}


\end{frontmatter}

%%%%%%%%%%%%%%%%%%%%%%%%%%%%%%%%%%%%%%%%%%%%%%%%%%%%%%%%%%%%%%%%%%%%%%%%%%%%%%
\pagebreak 

\section{Introduction}

\label{sec:intro}
$\quad$ \\

 If someone is infected with severe acute respiratory syndrome coronavirus 2 (SARS-CoV-2), the pathogen that causes COVID-19, how likely is that person to die of COVID-19?  This  simple question is surprisingly difficult to answer.  
 
 The ``case fatality rate'' (CFR) is a common measure that quantifies the mortality risk in a certain population, and is given by the ratio of deaths ($D$) over confirmed cases ($CC$) during a specific time period. However, because many COVID-19 cases are never diagnosed, the CFR almost certainly overestimates the true lethality of the virus.  Instead, the better answer is captured by the infection fatality rate (IFR) \citep{kobayashi2020communicating,wong2013case}.  The IFR, also a simple ratio, differentiates itself from the CFR by considering all cases, including the asymptomatic, undetected and misdiagnosed infections, in the denominator. For instance, if 20 individuals die of the disease in a population with 1,000 infections, then the IFR is 20 / 1000 = 0.02 = 2\%.
 %, whereas the CFR will be larger depending on how many cases have been diagnosed.
 % or have been assessed using diagnostics with limited accuracy
 
 Evidently, a key challenge in calculating the IFR is determining the true total number of cases. The total number of cases ($C$) is not known because not everyone is tested in the population ($P$).  A na\"{i}ve estimate of the IFR might take this into account by simply considering the number of tests ($T$) and estimating the number of cases as: $C \approx (CC/T) \times P$. However, diagnostic tests are often selectively initiated, such that tested individuals are not representative of the population at large.  
 
 %For this reason, it is helpful to consider the total number of confirmed cases (CC; which is used in the CFR) and the total number of unconfirmed cases.    A na\"{i}ve estimate of the IFR can then be obtained by simply considering the number of tests
 
 In most countries/jurisdictions, those with classic COVID-19 symptoms (e.g. fever, dry cough, loss of smell or taste) are much more likely to be tested than those without symptoms.  Due to this ``severity bias,'' the reported number of cases likely includes mostly people whose symptoms were severe enough to be tested and excludes the vast majority of those who are mildly- or asymptomatic. Even when testing is made equally available to all individuals, there is potential for bias if people who have reason to believe they are infected are more likely to volunteer to be tested (e.g., \citet{bendavid2020covid}).  We refer to the phenomenon whereby infected individuals are more likely to be tested than non-infected individuals,  as ``preferential testing''  (\cite{hauser2020estimation} and others use the term ``preferential ascertainment.'')

%Severity of disease is related to both the probability of being tested and the probability of death as the "classic" COVID-19 symptoms, which increase the likelihood of receiving a test, may occur in the latter stages of disease progression when antivirals and oxygen therapy are less effective. Inequities in access to care are reflected in access to COVID-19 testing.  In the US, racial and ethnic minorities and undocumented immigrants are less likely to be referred to testing than white Americans, irregardless of symptoms. 

If the degree of preferential testing in a particular sample is of known magnitude, bias adjustment can be achieved by appropriately altering the estimated rate of infection and its uncertainty interval.  However, the degree of preferential testing is typically unknown and likely highly variable across different jurisdictions.  An open question is  whether or not it is possible to reliably estimate the IFR without any specific information about the degree to which the data are biased by preferential testing (Q1).  And, if we have some samples for which testing is representative and others which are subject to some unknown bias from preferential testing, is it better to use only the representative data or to combine both kinds of data in a joint analysis (Q2)? In this paper, we address these two important questions by considering a Bayesian hierarchical model for estimation of the IFR.  We demonstrate with an application of the model to European data from seroprevalence studies and national official COVID-19 statistics.

%using a Bayesian hierarchical model that incorporates heterogeneity in preferential testing across many jurisdictions.

%and can make use of strong prior information on degree of preferential testing for some jurisdictions.

%from early data from China and Italy
Bayesian models have been previously used in similar situations.  For example, \citet{presanis2009severity} conduct Bayesian inference to estimate the severity of pandemic H1N1 influenza.  More recently, \citet{rinaldi2020empirical}, and \cite{hauser2020estimation} use Bayesian models for disease dynamics in order to estimate the severity of COVID-19.   To address the issue of preferential testing bias, \cite{hauser2020estimation} apply susceptible-exposed-infected-removed (SEIR) compartmental models to age-stratified data and, in order to establish parameter identifiability, assume that all cases of infected patients aged 80 years and older are confirmed cases.  The Bayesian model we propose is more general and allows one to obtain appropriate point and interval estimates for the IFR with varying degrees of prior knowledge about the magnitude of preferential testing and the distribution of other explanatory factors (e.g. age, healthcare capacity).

%I haven't read those other papers, but just trying to read between the lines of your statements here, would it be fair to say that they model disease dynamics and anchor identifiability by assuming that all cases of infected patients aged 80 years and older are confirmed cases, whereas we model snapshots of confirmed cases, which may be categorized by age or other variables.  Again, I am just guessing here.

This paper is structured as follows.  In Section 2, we introduce required notation, discuss distributional assumptions and review key issues of identifiability.  In Section 3, we formulate our Bayesian model and in Section 4, we describe how the model can be scaled for larger populations and can incorporate covariates.  In Section 5, we present a simulation study and in Section 6, we present an analysis of COVID-19 data from Europe.  We conclude in Section 7 with a return to the primary questions of interest (Q1 and Q2).

\section{Notation, distributions, and issues of (un)identifiability}
\label{sec:notation}
\subsection{Notation and distributions}
\label{sec:notation_sub}

 Suppose we have data from $K$ independent groups (i.e., countries or jurisdictions) from a certain fixed period of time.  For group $k$ = $1, \ldots, K$, let:
 
 \begin{itemize}
\item $P_{k}$ be the population size (i.e., the total number of individuals at risk of infection);
\item $T_{k}$ be the total number of people tested;
 \item $CC_{k}$ be the total number of confirmed cases resulting from the tests; and
\item $D_{k}$ be the total number of observed deaths attributed to infection.
\end{itemize}

\noindent We do not observe the following latent variables.  For the $k$-th group, let:

\begin{itemize}

\item $C_{k}$ be the total number of infected people (cases) in the population;
\item $IR_{k}$ be the true infection rate (proportion of  the population which is infected), which is the expected value of  $C_k/P_k$; and
\item $IFR_{k}$ be the true underlying infection fatality rate (IFR), which is the expected value of  $D_k/C_k$.
\end{itemize}

\noindent Therefore, we assume that: 
\begin{align}
 \label{eq:C_binom}
 C_{k} &\sim {Binom}(P_{k}, IR_{k}),     \quad \textrm{and:} \\
 \label{eq:D_binom}
D_{k}|C_{k} &\sim {Binom}( C_{k}, IFR_{k}),    
\end{align}
\noindent where, in the $k$-th group, the unknown number of infections, $C_{k}$, and the known number of deaths, $D_{k}$, each follow a binomial distribution.  Note that there are latent variables on both the left hand side and the right hand side of (\ref{eq:C_binom}).

For each group, $CC_{k}$ is recorded, instead of $C_{k}$.  Even in the absence of preferential testing, $CC_{k}$ will be smaller than $C_{k}$ because not everyone is tested.  The goal is to draw inference on the relationship between the number of deaths, $D$, and the number of cases, $C$, having only data on $D$, $CC$, $P$, and $T$.  This is particularly challenging since the number of confirmed cases in each group may be subject to an unknown degree of preferential testing.   
   
   In the absence of any preferential testing, if one assumes that the population sizes are finite, then the number of confirmed cases will follow a hyper-geometric distribution \citep{prochaska2018discrete}.  The hyper-geometric distribution describes the probability of $CC_{k}$ confirmed cases amongst $T_{k}$ tests (without any individuals being tested more than once), from a finite population of size $P_{k}$ that contains exactly $C_{k}$ cases.  Wallenius' \emph{non-central} hyper-geometric is a generalization of the hyper-geometric distribution whereby testing is potentially biased with either cases or non-cases more likely to be tested \citep{fog2008sampling}.  We therefore consider the distribution of $CC_{k}|C_{k}$ as following a {non-central} hyper-geometric (NCHG) distribution:
   \begin{align}
   CC_{k} |  C_{k} &\sim {NCHyperGeo}(C_{k}, P_{k}-C_{k}, T_{k}, \phi_{k}),
   \label{eq:CC_NCHG}
\end{align}
    \noindent where the degree of preferential testing corresponds to the $\phi_{k}$ non-centrality parameter \textcolor{black}{(see Appendix (Section \ref{sec:noncentralhyper}) for details about the NCHG distribution)}.  When $\phi_{k}>1$, cases (i.e., infected individuals) are more likely to be tested than non-cases (i.e., non-infected individuals);  when $\phi_{k}<1$, cases are less likely to be tested than non-cases.  When $\phi_{k}=1$, we have that the probability of being tested is equal for both cases and non-cases, and the NCHG distribution reduces to the standard hyper-geometric distribution.  In this parameterization, the $\phi_{k}$ parameter can be interpreted as an odds ratio: the odds of a case being tested vs. the odds of a non-case being tested.   

\textcolor{black}{The distribution of the confirmed cases depends on the actual infection rate ($C/P$) and the testing rate ($T/P$), but does not depend on the infection fatality rate ($D/C$).  In other words, we assume that the conditional distribution of $(CC|C,T,P,D)$ is identical to the conditional distribution of $(CC|C,T,P)$.  This assumption is similar to the assumption of ``non-differential'' exposure misclassification in measurement error models and may or may not be realistic; see \cite{de2018bias}.  If across the $K$ different groups, those groups with higher $\phi_{k}$ values also tend to have higher $IFR_{k}$ values, then one will inevitably obtain biased estimates because the $IFR_{k}$ and $\phi_{k}$ are considered \textit{a priori} independent.  The same logic applies to the $IR_{k}$ and $IFR_{k}$ which are also \textit{a priori} independent.}  

\textcolor{black}{Also note that the members of set $D_{k}$ are not a subset of the members of set $CC_{k}$.  While $D_{k}$ is a subset of $C_{k}$,  and $CC_{k}$ is a subset of $C_{k} \cap T_{k}$,  $D_{k}$ is not necessarily a subset of $CC_{k}$. For example, in the seroprevalence study data for Luxembourg which we consider in Section \ref{sec:eur} (see Table \ref{tab:eur}, row 3), we have $CC_{k} = 23$ confirmed cases out of $T_{k} = 1,214$ tests.  There are $D_{k} = 93$ deaths out of a population of $P_{k} = 615,729$.  Evidently,   $D_{k}$ is not a subset of $CC_{k}$.  Furthermore, the assumption that $\phi_{k} = 1$ for this Luxembourg data implies that the 1,214 tested individuals were not any more or less likely to be infected than those in the general population.  However, note that there is no requirement that the tested individuals have the same risk of death as those in the general population. To be clear, no distributional assumptions will be violated if, within the $k$-th group, individuals with a higher probability of death (e.g., the elderly) are more likely to be tested than those with a lower probability of death (e.g., young, healthy individuals).}

 \subsection{Partial identifiability}
 \label{sec:partialid}
  Given the assumptions detailed above, for each of the $K$ groups, there are three unknown parameters (latent states), $IR_{k}$, $IFR_{k}$ and $\phi_{k}$, that must be estimated for every two observed quantities ($D_{k}/P_{k}$ and $CC_{k}/T_{k}$).  This indicates that a unique solution will not be attainable without additional external data or prior information.  
 
 The problem at hand is sufficiently rich and complex that forming intuition about the information-content of the data is challenging.    In the Appendix (Section \ref{sec:unident}), we consider, in depth, an asymptotic argument for \emph{partial identifiability}.  We determine that, depending on the range and heterogeneity in the degree of preferential testing across groups, the data can contribute substantial information about the infection fatality rate.  Data from any single group may only be  weakly informative about the IFR, in the sense that only lower and upper bounds for the IFR are estimable.  However, we show that in some circumstances there is very considerable sharpening of information when these bounds are combined across groups, provided it is {\em a priori} plausible that the IFR heterogeneity across groups is modest.

\section{A Bayesian model for small-$P$ data} 

\label{sec:smallP}
$\quad$ \\

We describe a Bayesian model which assumes standard {\color{black}Gaussian random-effects allowing both the infection rate (IR) and infection fatality rate (IFR) to vary between groups.  Bayesian models are known to work well for dealing with partially identifiable models; see \citet{gustafson2010bayesian}.    Consider the following random-effects model: }
\begin{align}
\label{eq:IFR}  
   \operatorname{g}(IFR_{k}) &\sim \mathcal{N}(\theta, \tau^{2}), \quad  \textrm{and}  \\
\label{eq:IR}    
   \operatorname{g}(IR_{k}) &\sim \mathcal{N}(\beta, \sigma^{2}),
\end{align}
%$\operatorname{cloglog}(IFR_{k}) \sim \mathcal{N}(\theta, \tau^{2})$, and  %\label{eq:IFR}   \\ 
%$\operatorname{cloglog}(IR_{k}) \sim \mathcal{N}(\beta, \sigma^{2})$, %\label{eq:IR} 
%\end{align} 
%
\noindent for $k$ = $1, \ldots, K$, where $\theta$ is the parameter of primary interest, $\tau^{2}$ represents between group IFR heterogeneity, $\beta$ represents the mean g(infection rate), $\sigma^{2}$ describes the variability in infection rates across the $K$ groups, and g() is a given link function.   {Note that, alternatively, a simpler fixed-effects version of the model arises by setting $\tau=0$ such that  $  \operatorname{g}(IFR_{k}) = \theta,$ for $k=1,\ldots,K$.}

 \textcolor{black}{ We will adopt the complimentary log-log link function (cloglog) for g(), though there are other sensible choices including the logit and probit functions.  Our choice of the cloglog function facilitated the creation of parameter-transformed samplers for efficient sampling (see Appendix \ref{sec:nimble}).} 

Putting together the assumptions for $p(D_{k} | IFR_{k}, C_{k})$, $p(CC_{k} | T_{k}, P_{k}, C_{k},  \phi_{k})$ and $ p(C_{k}|P_{k}, IR_{k})$ defined in Section \ref{sec:notation_sub} along with prior distributions, Bayes' Law takes the form:
%
%A Bayesian model begins by specifying a joint probability distribution.  For our unknown parameters of interest ($\theta$, $\tau^{2}$, $\beta$, $\sigma^{2}$), latent variables ($C_{k}$, $IFR_{k}$, $IR_{k}$, and $\phi_{k}$,  for $k$ = 1,$\ldots, K$), and aggregate data from $K$ sources (we require data = $\{P_{k}$, $T_{k}$, $CC_{k}$, and $D_{k}\}$, for $k$ = 1,$\ldots, K$), Bayes theorem states that:
%
\begin{align}
\label{eq:bayesAD}
    p( (\theta, \tau^{2},  \beta, \sigma^{2}, \textrm{C}, \textrm{IFR}, \textrm{IR}, \phi)&  | \textrm{data}) \propto p( \textrm{data} |\theta, \tau^{2},  \beta, \sigma^{2}, \textrm{C},  \textrm{IFR}, \textrm{IR},  \phi) \\ \nonumber
 &   \quad  \quad\quad  \quad \quad   \times p(\theta, \tau^{2},  \beta, \sigma^{2}, \textrm{C}, \textrm{IFR}, \textrm{IR}, \phi)   \\ \nonumber
= \Big(\prod_{k=1}^{K}p(D_{k}  | IFR_{k}, C_{k}) &p(CC_{k} | T_{k}, P_{k}, C_{k},  \phi_{k}) p(C_{k}|P_{k}, IR_{k}) p(IFR_{k}   | \theta, \tau^{2}) p(IR_{k} | \beta, \sigma^{2})  \Big)  \\ \nonumber
   &   \quad  \quad \quad \quad \times p(\theta) p( \tau^{2} ) p(\beta) p( \sigma^{2} )  p(\phi).
\label{eq:bayesAD2}
\end{align}
%
  % We have that $p(D_{k} |  IFR_{k}, C_{k}) $ is defined according to a  binomial distribution as stated in  (\ref{eq:D_binom}),  that $p(CC_{k} | T_{k}, P_{k}, C_{k},  \phi_{k})$ is defined according to  (\ref{eq:CC_NCHG}), and that $p(C_{k}|P_{k}, IR_{k}) $ is defined by (\ref{eq:C_binom}).  We also have that $ p(IFR_{k}   | \theta, \tau^{2})$ and $p(IR_{k} | \beta, \sigma^{2})$ are defined according to (\ref{eq:IFR}) and (\ref{eq:IR}) respectively.  
  
  %We have that $p(D_{k} |  IFR_{k}, C_{k}) $, $p(CC_{k} | T_{k}, P_{k}, C_{k},  \phi_{k})$, and $p(C_{k}|P_{k}, IR_{k}) $ are defined by the distributions assumed in Section 2.1.  
 
 %Many warn against uniform priors, arguing that they are not truly uninformative, and may give too much prior weight to unreasonably large values.  As an alternative to ``uninformative'' priors, weakly-informative priors, are often recommended; see \cite{gelman2006prior}. 
 
 % berger2013statistical, burke2018bayesian, lambert2005vague
 
  We are left to define prior distributions for the unknown parameters: $\theta$, $\tau^{2}$, $\beta$, $\sigma^{2}$, and $\phi$.   Our strategy for priors on IR and IFR is to assume uninformative priors for the mean of IFR and of IR and for the variance of IR, but a strongly informative prior favouring small values for the variance of IFR.  This strategy reflects the assumption that the infection fatality rate varies across jurisdictions much less than the infection rate itself (especially after accounting for population level sources of heterogeneity; see Section \ref{sec:covariates}).  Uniform and half-Normal priors are set accordingly: $ \operatorname{g}^{-1}(\theta) \sim Uniform(0, 1);  $   $\operatorname{g}^{-1}(\beta) \sim Uniform(0, 1);  $  $  \sigma \sim \textrm{half-}\mathcal{N}(0, 1)$ and   $  \tau \sim \textrm{half-}\mathcal{N}(0, \eta^{2})$ , where $\eta=0.1.$

 The only remaining component is $p(\phi)$.   Our strategy for a prior on the degree of preferential testing is to assume that cases are more likely to be tested than non-cases (i.e., $\phi_{k} > 1$), that all values of $\phi_{k}$ are equally likely across jurisdictions, and that there is an upper bound, $1+\gamma$, on the degree of preferentiality.  For the upper bound parameter, $\gamma$, we assume an exponential prior, such that:
 
 $\phi_{k}|\gamma  \sim Uniform(1, 1 + \gamma), \quad \textrm{for $k$ = 1,}\ldots,K$; $\quad $ and  $\quad  \gamma \sim {Exp}(\lambda).$ 
 
 \noindent \textcolor{black}{We therefore assume that the uniform range of possible values for $\phi_{k}$ is itself unknown.  This hierarchy allows one to specify a very ``weakly informative'' prior for the degree of preferential testing.  For instance, setting $\lambda=0.05$ implies that, \emph{a priori}, a reasonable value for the $\phi_{k}$ odds ratio is about 6 (infected individuals are about 6 times more likely to be tested than those uninfected) and could range anywhere from about 3 to 14.  (When $\lambda=0.05$, the median of the unconditional distribution for $\phi_{k}$ is 6.4, with a wide interquartile range of 2.7 to 14.3.)}
   
 In some scenarios, we might have some groups for which $\phi_{k}$ is known and equal to 1 (i.e., have data from some samples where testing is known to be truly random).  Without loss of generality, suppose this subset is the first $k^{'}$ studies, such that for $k= 1, \ldots, k^{'}$, we have $\phi_{k}=1$.  We will use this approach in the European data analysis (Section \ref{sec:eur}), in which we assume $\phi_{k}$ is known and equal to 1 for data from representative seroprevalence studies.
 
%  Thus, in a situation where testing is known to be random for all groups, $k^{'}=K$.  
 %and $p(\phi_{k})=1$ in (\ref{eq:bayesAD})
 %Alternatively, when preferential testing cannot be ruled out for certain groups, the unknown $\phi_{k}$ values for $k>k^{'}$ are assumed greater than 1, and defined by the uniform prior above.

We must emphasize that the performance of any Bayesian estimator will depend on the choice of priors and that this choice can substantially influence the posterior when few data are available \citep{berger2013statistical, lambert2005vague}.  The priors described here represent a scenario where there is little to no \emph{a priori} knowledge about the $\theta$, $\beta$, and $\phi$ model parameters.  Inference would no doubt be improved should more informative priors be specified based on probable values for each of these parameters.  We will consider the impact of priors in the simulation study in Section \ref{sec:simstudy}, where we look to different values for $\lambda$ \textcolor{black}{and $\eta$}.  
 
{
 We must also emphasize that, due to the partial identifiability issues (Section \ref{sec:partialid}), a delicate trade-off may exist between the priors for the $\tau$ and $\phi$ parameters.  For instance, if large values of $\tau$ are made \emph{a priori} plausible (i.e., if $\eta$ is large), then the posterior estimates of the $\phi$ parameters may be driven downwards towards 1 (due to the $ \gamma \sim {Exp}(\lambda)$ prior).  A relatively homogeneous across-group IFR can be central to identifiability and, as such, the aforementioned ``fixed-effects'' version of the model (essentially equivalent to fixing $\tau=0$) may be more feasible in situations when identification is particularly challenging (e.g., when $k^{'}=0$, and/or when there is very little prior knowledge about the $\theta$, $\beta$, and $\phi$ model parameters).  \textcolor{black}{On the other hand, in situations when identification is less of a concern (e.g., when $k^{'}$ is relatively large relative to $K$ and/or when there is substantial and reliable prior information), setting \emph{a priori} limitations on $\tau$ may be detrimental if the true heterogeneity in infection fatality rates across groups is high and meaningful.}}

% \begin{itemize}
%     \item MCMC is used to fit the model. \cite{auzenbergs2019desirable} describe using the BUGS software for infectious disease models.
 
%    \item \cite{chatzilena2019contemporary} describe using Stan software for infectious disease models.
 
 %   \item Great resource for priors with JAGS https://onlinelibrary.wiley.com/doi/pdf/10.1002/9781119942412.app1

%\end{itemize}

% a hyper-geometric distribution is asymptotically equivalent to a binomial distribution.
 %particular {parameterization} of does not emerge from the limit of the non-central hyper-geometric distribution.

\section{A Bayesian model for large-$P$ data}
\subsection{Distributional approximations}
\label{sec:largeP}

\noindent When populations are sufficiently large, two simplifications to the model are desirable. First, we will replace the NCHG distribution with a binomial distribution as follows:
\begin{equation}
CC_{k} \sim {Binom}(T_{k}, 1-(1-{C_{k}}/{P_{k}})^{\phi_{k}}) 
\label{eq:ccmid}
\end{equation} 
%
%in (\ref{eq:CC_NCHG})
%hannan1963normal
\noindent \textcolor{black}{for $k$ = $1, \ldots, K$.  This simplification\footnote{Recall that a hyper-geometric distribution is asymptotically equivalent to a binomial distribution and while this particular binomial {parameterization} does not emerge from the limit of the NCHG distribution, it is a reasonable approximation.  We could have alternatively substituted the NCHG distribution  with the known Gaussian asymptotic approximation to the NCHG \citep{ stevens1951mean}.  However, the  Gaussian  approximation requires solving quadratic equations and therefore might not actually make things simpler; see \citet{sahai1995statistics}.} alleviates the need for writing custom samplers for the NCHG distribution for certain MCMC software (e.g., Stan, nimble) and also provides additional familiarity to researchers who may not be accustomed to working with the NCHG distribution.  Secondly, we can dispense with the need to sample the $C_{k}$ latent variables by replacing the above distribution for $CC_{k}$ with:}
\begin{equation}
CC_{k} \sim {Binom}(T_{k}, 1-(1-IR_{k})^{\phi_{k}}).
\label{eq:ccfinal}
\end{equation} 
\textcolor{black}{\noindent For any sufficiently large $P_{k}$, this simplification will make little to no difference.  Then, since the distributions of $C_{k}$ and $D_{k}|C_{k}$ are both binomials (see (\ref{eq:C_binom}) and (\ref{eq:D_binom})), we have that unconditionally: }
\begin{eqnarray}
 \label{eq:D_uncond}
D_{k} &\sim {Binom}(P_{k}, IFR_{k} \times IR_{k}).
\end{eqnarray}

Note that in (\ref{eq:ccmid}) and (\ref{eq:ccfinal}) above, the $\phi_{k}$ parameter no longer corresponds to an odds ratio, yet the interpretation is similar.
Starting from (\ref{eq:ccfinal}), the odds ratio (OR) 
describing the association between testing status and infection status is
\begin{eqnarray*}
\log (OR) &=&
\log(1-(1-IR)^\phi) - \phi \times \log(1-IR) - \log(IR) + \log(1-IR).
\end{eqnarray*}
For fixed $IR$, approximating this with a Taylor series in $\log(\phi)$, about zero, gives: $\log (OR) \approx c_{IR} \log(\phi),$ where $c_{IR} = -\log(1-IR)/IR$.  Note that $c_{IR} \rightarrow 1$ as $IR \rightarrow 0$.  Therefore, in the rare-infection realm, $\phi$ is indeed approximately the odds ratio for testing and infection status.

 %For our unknown parameters of interest ($\theta$, $\tau^{2}$, $\beta$, $\sigma^{2}$), latent variables ($IFR_{k}$, $IR_{k}$, and $\phi_{k}$,  for $k$ = 1,$\ldots, K$), and aggregate data from $K$ sources (we require data = $\{P_{k}$, $T_{k}$, $CC_{k}$, and $D_{k}\}$, for $k$ = 1,$\ldots, K$), Bayes' theorem states that:
%
%\begin{align}
%    p( (\theta, \tau^{2}, & \beta, \sigma^{2}, \textrm{IFR}, \textrm{IR}, \phi)  | \textrm{data}) \propto p( \textrm{data} |\theta, \tau^{2},  \beta, \sigma^{2}, \textrm{IFR}, \textrm{IR},  \phi) \\ \nonumber
% & \quad   \times p(\theta, \tau^{2},  \beta, \sigma^{2},\textrm{IFR}, \textrm{IR}, \phi)  \\ \nonumber
%= \Big(\prod_{k=1}^{K}p(D_{k} & | IFR_{k}, IR_{k}, P_{k}) p(CC_{k} | T_{k}, IR_{k},  \phi_{k})  p(IFR_{k}   | \theta, \tau^{2}) p(IR_{k} | \beta, \sigma^{2})  \Big)  \\ \nonumber
%   & \quad \quad \times p(\theta) p( \tau^{2} ) p(\beta) p( \sigma^{2} ) \prod_{k=1}^{K}  p(\phi_{k}),
%\label{eq:bayesAD_large}
%\end{align}
%
%\noindent where $p(D_{k}  | IFR_{k}, IR_{k}, P_{k})$ and $p(CC_{k} | T_{k}, IR_{k}, \phi_{k})$ are defined by binomial distributions as detailed above, and priors are defined as in Section \ref{sec:21}.  

\subsection{Including group-level covariates}
\label{sec:covariates}

\textcolor{black}{The proposed model can be expanded to include factors of interest specified as covariates at the group level, resembling what is commonly done in a meta-regression analysis \citep{thompson2002should}. Covariates included for analysis might be metrics that are correlated with the probability of infection, with the probability of being tested, with the accuracy of the test, and/or with the probability of dying from infection.}

For instance, suppose that $X_{[1]k},\ldots, X_{[h]k}$ are $h$ different group-level covariates that explain the $k$-th group's infection rate, and that $Z_{[1]k},\ldots, Z_{[q]k}$ are $q$ different covariates that explain the $k$-th group's IFR.  Then these can be incorporated as follows:
\begin{align}
\operatorname{g}(IR_{k}) &\sim \mathcal{N}(\beta + \beta_{1}X_{[1]k} + \hdots + \beta_{h}X_{[h]k}, \sigma^{2}), \\
\operatorname{g}(IFR_{k}) &\sim \mathcal{N}(\theta + \theta_{1}Z_{[1]k} + \hdots + \theta_{q}Z_{[q]k}, \tau^{2}).
\end{align}
Age is a key factor for explaining the probability of COVID-19-related death \citep{o2020age}.  One might therefore consider median age of each group as a predictor for the IFR, or perform analyses that are stratified by different age groups \citep{onder2020case}.  The latter strategy has, for instance, been recommended to make accurate predictions for respiratory infections \citep{pellis2020systematic}.  With regards to the infection rate, time since first reported infection, or time between first reported infection and the imposition of social distancing measures might be predictive \citep{anderson2020will}.

%; see also \cite{cantsmell, Walkerm2808}.

\subsection{MCMC}
\label{sec:mcmc}
For the large-$P$ model, Markov chain Monte Carlo (MCMC) mixing can be slow because different combinations of $\phik$, $\cIRk$ and $\cIFRk$ can yield similar model probabilities.  This is related to the identifiability issues discussed in the Appendix (Section \ref{sec:unident}).  Standard Gibbs sampling (e.g., as implemented with JAGS \citep{kruschke2014doing}) will be inefficient in many situations.  To improve mixing and reduce computational time, we wrote the model in the nimble package \citep{de2017programming}, which supports an extension of the modeling language used in JAGS and makes it easy to configure samplers and provide new samplers.  Details of the MCMC implementation for nimble are presented in the Appendix (Section \ref{sec:nimble}).  We also implemented the large-$P$ model in the popular Stan package which employs Hamiltonian MCMC algorithms \citep{carpenter2017stan}.
%All code is publicly available online at XXX.

%\subsection{MCMC details}

%Mixing is an issue...  

%Since several parameters are correlated we can improve mixing by...

%We were also able to transform the parameters... 

%Initial values...

%Nimble is a wonderful tool for fitting these types of models; see \citet{de2017programming}.

\section{Simulation study}
\label{sec:simstudy}
\subsection{Design}

We conducted a simulation study in order to better understand the operating characteristics of the proposed model.  \textcolor{black}{Specifically, we wished to evaluate the frequentist coverage of the credible interval for $\theta$ and investigate the impact of choosing different priors.}

As emphasized in \cite{gustafson2009interval}, the average frequentist coverage of a Bayesian credible interval, taken with respect to the prior distribution over the parameter space, will equal the nominal coverage.  This mathematical property is unaffected by the lack of identification.  However, the variability of coverage across the parameter space is difficult to anticipate and could be highly affected by the choice of prior.  For example, we might expect that, in the absence of preferential testing (i.e., when $\gamma=0$), coverage will be lower than the nominal rate.  However, if this is the case, coverage will need to be higher than the nominal rate when $\gamma>0$, so that the ``average'' coverage (taken with respect to the prior distribution over the parameter space) is nominal overall.

\textcolor{black}{ We simulated datasets with $K=20$ and $k^{'}=8$.  For $k=1, \ldots, 8$, population sizes were obtained from a $NegBin(20000, 1)$ distribution with a mean of 20,000 and for $k=9, \ldots, 20$, population sizes were obtained from a $NegBin(200000, 1)$ distribution.  Parameter values were as follows: $\theta = cloglog(0.02) = -3.90$, $\beta=cloglog(0.20) = -1.50$, $\tau^{2}=0.005$ and $\sigma^{2}=0.25$.  The testing rate for each population was obtained from a $Uniform(0.01, 0.10)$ distribution so that the proportion of tested individuals in each population ranged from 1\%  to 10\%.  We considered eight values of interest for $\gamma$:  0,  0.5, 1,  2, 4,   12,  32,   64 (for simulation); and three different values of interest for both $\lambda$ and $\eta$:  0.05, 0.1, and 0.5 (for estimation).  The number of confirmed cases ($CC_{k}$) were simulated from Wallenius' NCHG distribution as detailed in Section \ref{sec:notation_sub}.  The 12 ``unknown'' $\phi_{k}$ values, for $k$ = $9,\ldots,20$, were simulated from a $Uniform(1, \gamma+1)$ distribution.  Note that with high $\gamma$ levels, the vast majority of tests will be positive (when $\gamma=32$, positivity is about 72\%; when $\gamma=64$, positivity is about 81\%).  }
   
 We fit three models to each unique dataset: $M{1}$, $M{2}$, and $M{3}$.  All three models follow the same large-$P$ framework detailed in Section \ref{sec:largeP}, but each considers a different subset of the data:
 
 \begin{itemize}
 \item{ The $M{1}$ model uses only data from the groups for which $\phi_k$ is unknown, i.e., $\{P_{k}$, $T_{k}$, $CC_{k}$, and $D_{k}\}$ for $k$ = $9,\ldots,20$ ($k^{'}=0$, and $K=12$);}
 
  \item{The $M{2}$ model considers the data from all 20 groups,  i.e., $\{P_{k}$, $T_{k}$, $CC_{k}$, and $D_{k}\}$ for $k$ = $1,\ldots,20$ ($k^{'}=8$, and $K=20$); and}
  
  \item{The $M{3}$ model uses only data from the groups for which $\phi_k$ is known and equal to 1,  i.e., $\{P_{k}$, $T_{k}$, $CC_{k}$, and $D_{k}\}$ for $k$ = $1,\ldots,8$ ($k^{'}=8$, and $K=8$). }

 \end{itemize}
 
\noindent To be clear, the $M{2}$ and $M{3}$ models make the assumption of (correctly) known $\phi_{k}=1$ for $k = 1,\ldots,8$.
  
\textcolor{black}{We simulated 1,100 unique datasets (i.e., 1,100 unique sets of values for $\{P_{k}, T_{k}, D_{k}, CC_{k,\gamma}\}$, for $k=1,\ldots,K$, and $\gamma=\{0,  0.5, 1,  2, 4,  12,  32,   64\}$) and, for each dataset, fit the three different models.  (See  Table \ref{tab:ex1} in the Appendix for an example of a ``single'' unique dataset.)  We specifically chose to conduct 1,100 simulation runs so as to keep computing time within a reasonable limit while also reducing the amount of Monte Carlo standard error (MCSE) to a reasonably small amount.  For looking at coverage with $1-\alpha=0.90$, MCSE will be approximately $\sqrt{0.90(1-0.90)/1100} \approx 0.009$; see \cite{morris2019using}.}

\textcolor{black}{For each unique dataset, the $M{1}$ and $M{2}$ models were fit 72 times ($= 3 \times 3 \times 8$): with $\lambda$ assuming one of the 3 values of interest, with $\eta$ assuming one of the 3 values of interest, and with one of the 8 different sets of $CC_{k}$ numbers (for $k$ = $9,\ldots,20$) corresponding to the nine $\gamma$ values of interest.  The  $M{3}$ model was fit 3 times for each unique dataset: with $\eta$ assuming one of the three values of interest.  For each model fit, we recorded the the posterior median estimate of $icloglog(\theta)$, the width of the 90\% highest posterior density (HPD) CI for $\theta$, and whether or not the 90\% HPD CI contained the target value of $cloglog(0.02)=-3.90$.}
   
\textcolor{black}{  For each simulation scenario, we used Stan to obtain a minimum of $N_{MC}=18,000$ MCMC draws from the posterior (a total from 3 independent chains, with 20\% burn-in, and thinning of 5).  We recorded the Gelman-Rubin test statistic,  $\hat{R}$ \citep{gelman1992inference, brooks1998general} and if this statistic was $ \hat{R}>1.05$, the MCMC sampling was discarded and was restarted anew with twice the number of MCMC draws, up to a maximum of $N_{MC}=288,000$.  If, even after 4 re-starts, with $N_{MC}=288,000$, we obtained $\hat{R}>1.05$, a convergence/mixing failure was recorded and the result was simply discarded.}
 
\subsection{Results}
 \label{sec:simresults}
 
\begin{sidewaysfigure}[p]
    \centering
    \includegraphics[width=19.5cm]{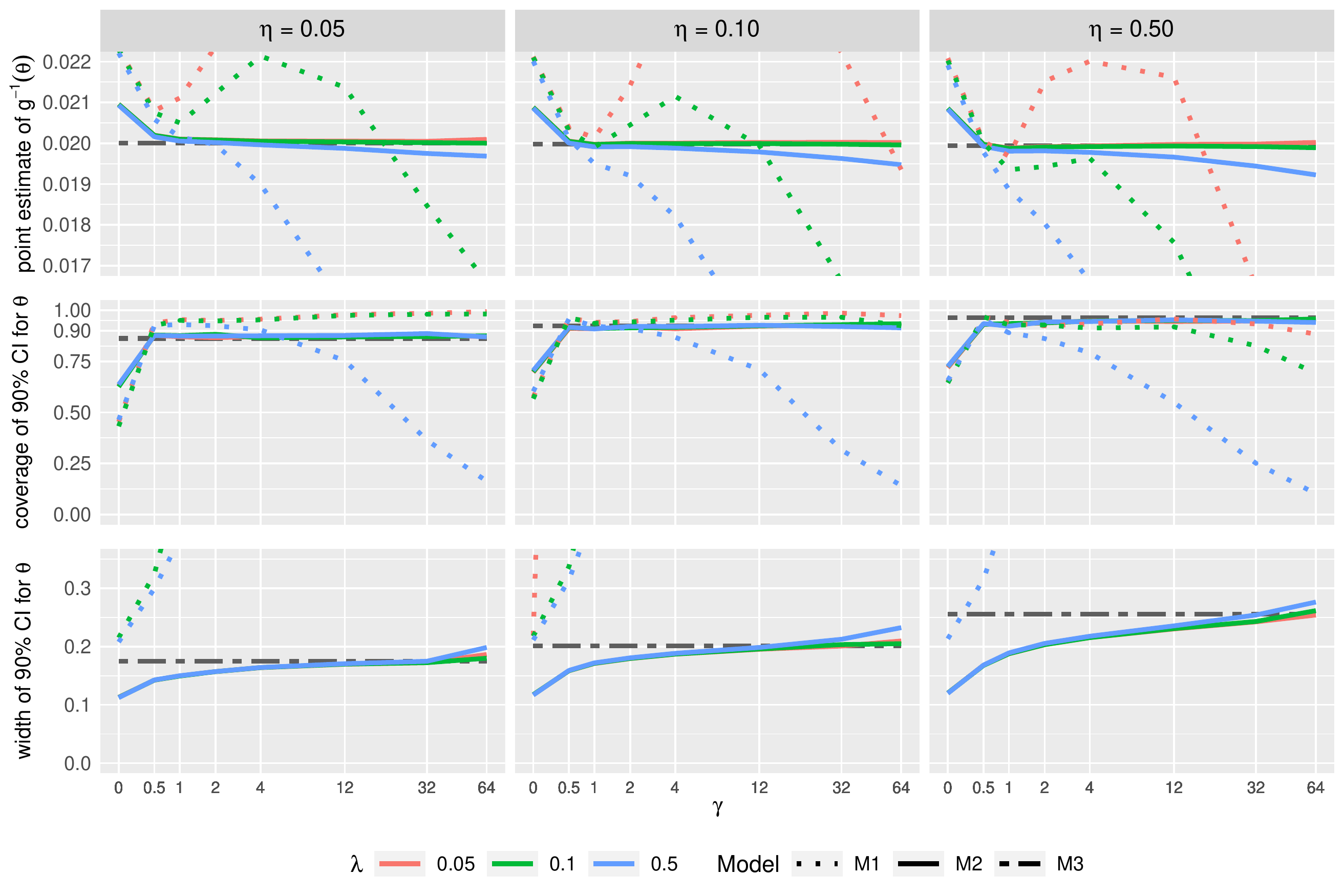}
    \caption{Results from the simulation study.  The top row plots average point estimate obtained for $icloglog(\theta)$, the middle row plots the coverage of the 90\% HPD CI; and the bottom row plots the average width of the 90\% HPD CI.  Each column of panels corresponds to a different level of $\eta$.  To be clear, each different value for $\gamma$ corresponds to a different upper bound on the degree of preferential testing in the simulated data. Different values for $\lambda$ and $\eta$ correspond to different prior specifications.}
    \label{fig:sim_res2}
\end{sidewaysfigure}
 
\textcolor{black}{Figure \ref{fig:sim_res2} plots the simulation study results.  The $M{3}$ model, which only  considers data from those groups where testing is known to be representative/random, appears to obtain a average point estimate for $icloglog(\theta)$ of approximately 0.02 as desired for all three values of $\eta$.  In contrast, the $M{1}$ model, which only considers data from those groups where the degree of preferential testing is unknown, obtains average point estimates for $icloglog(\theta)$ far above and far below the target value of 0.02 depending on $\eta$, $\lambda$, and $\gamma$.  (Note that many results for $M{1}$ are so large/small that they are outside the limits of the plot). The $M{2}$ model, which makes use of all the data, obtains point estimates of approximately 0.02 for all positive values of $\gamma$, when $\lambda$ is sufficiently small (i.e., $\lambda \le 0.1$) for all three values of $\eta$.  When $\lambda=0.5$, the $M{2}$ model tends to underestimate $icloglog(\theta)$ when $\eta$ and/or $\gamma$ are large.}

\textcolor{black}{Coverage for models $M{2}$ and $M{3}$ appears to be highly dependent on $\eta$.  With $\eta = 0.1$, the $M{2}$ and $M{3}$ models obtain coverage of approximately 90\% as desired for all $\gamma>0$ values considered.  With $\eta = {0.05}$, coverage is ever so slightly less than the desired 90\% level and when $\eta = {0.5}$, coverage is higher than the desired 90\% level.  The results from the $M{1}$ model show that, for small values of $\lambda$ and $\eta$ (i.e., for $\lambda \le 0.5$ and $\eta < 0.5$) , coverage is at or above 90\% for the entire range of $\gamma$ values.  This suggests that appropriate coverage may be achievable even when $k^{'}=0$ and when in the presence of a substantial and unknown amount of preferential testing.}
 
\textcolor{black}{ The credible interval width results from the $M{2}$ and $M{3}$ models indicates that, for a wide range of $\gamma$ values, the $M{2}$ model (which makes use of all the data) is preferable to the $M{3}$ model (which uses data only from those groups where testing is known to be representative/random).  However, there is a limit to the ``added value'' that the ``non-representative'' data provide.  For example, for $\gamma>12$ and $\eta=0.1$, $M{3}$ intervals are narrower compared to $M{2}$ intervals (for all values of $\lambda$).}

\textcolor{black}{Overall, the interval width is much much narrower for $M{2}$ relative to $M{1}$.  This confirms that the $k^{'}=8$ representative samples are very valuable for reducing the uncertainty around $\theta$.  (Note that most credible interval width results for $M{1}$ are so large that they are outside the limits of the plot).  With regards to the COVID-19 pandemic, this emphasizes the importance of conducting some amount of ``unbiased testing'' even if the sample sizes are relatively small; see \cite{Cochran2020}. }  

\textcolor{black}{Finally, note that, if $k^{'}=0$, mixing can be problematic if $\lambda$ is small.  Indeed, for the $M{1}$ model, convergence/mixing failures occurred in about 1\% of simulation runs when $\lambda=0.05$, and occurred in less than 0.004\% of simulation runs when $\lambda>0.05$.  With very small $\lambda$ values (e.g., $\lambda<0.001$), we suspect that convergence may simply be impossible.  This is no doubt due to the identifiability issues discussed in Section  \ref{sec:partialid} and in the Appendix (Section \ref{sec:unident}).  If $k^{'}=0$, the model benefits greatly (in terms of mixing and identifiability) from specifying more informative priors.}

\section{Application-  IFR of COVID-19 in Europe}
\label{sec:eur}

$\quad$ \\

Reducing uncertainty around the severity of COVID-19 was of great importance to policy makers and the public during the early stages of the pandemic and continues to be a top priority \citep{iao2020, marc2020}.  Comparisons between the COVID-19 and seasonal influenza IFRs impacted the timing and degree of social distancing measures and highlighted the need for more accurate estimates for the severity of both viruses \citep{faust2020}. A lack of clarity means that policy makers are unsure if cross-population differences are related to clinically relevant heterogeneity (i.e., due to large $\tau$) or to spurious heterogeneity driven by testing and reporting biases (i.e., due to large $\gamma$).  

We demonstrate how the proposed model could be used to estimate the IFR of COVID-19 in Europe \textcolor{black}{during the spring of 2020}.  Note that the main purpose of this analysis is to demonstrate the feasibility of the proposed model.  As such, we keep things relatively simple.  For instance, we only consider countries belonging to the EU/EEA (European Economic Area), the United Kingdom, and Switzerland, as these could be considered a reasonably homogeneous group.  However, we exclude Belgium since, uniquely, the country counts all suspect deaths in nursing homes as COVID-19 deaths \citep{bbcbelgium}. %\footnote{as reported in https://www.bbc.com/news/world-europe-52491210}

We selected $k^{'}=5$ studies for which we assume there is no preferential testing.  To do so, we considered all European seroprevalence studies reporting an IR estimate (along with a 95\% confidence/credible interval) listed in the systematic review by \cite{ioannidis2020infection}.  From these, we selected only those studies that claimed to achieve a representative or random sample from their study population.

%A number of seroprevalence studies have attempted to obtain truly representative samples.  Following a literature review, we identified $k^{'}=5$ studies for which we could reasonably assume an absence of preferential testing (i.e., we assume that $\phi_{k}=1$, for $k=1,\ldots,5$).

%These $k^{'}=5$ studies were selected by first considering all European seroprevalence studies reporting an IR estimate (along with a 95\% confidence/credible interval) listed in the systematic review by \cite{ioannidis2020infection}.  We then selected only those studies that purported to achieve a ``representative sample'' from the target population.

It is important to note that the seroprevalence studies were conducted amongst populations which were particularly hard hit by infection.  The result is that these populations are not necessarily representative of the overall European population.  It is unclear how this  might impact our model estimates.  Also, while some of the seroprevalence studies report the exact number of tests conducted ($T$) and the number of confirmed cases recorded ($CC$), to obtain estimates for the infection rate, there are numerous adjustments (e.g., adjusting for testing sensitivity and specificity).  Rather than work with the raw $T_{k}$, and $CC_{k}$ numbers published in the seroprevalence studies, we calculate effective data values for $CC_{k}$ and $T_{k}$ based on a binomial distribution that corresponds to the reported 95\% CI for the IR.  By ``inverting binomial confidence intervals'' in this way, we are able to properly use the adjusted numbers for each of the five seroprevalence studies.  This is a similar approach to the strategy employed by \cite{Kummerer2020} who assume that the IR follows a Beta distribution with parameters chosen to match the 95\% CI published in \cite{streeck2020infection}.  In the Appendix (Section \ref{sec:serodetails}), we go over the seroprevalence study data in detail.

We obtained national official COVID-19 statistics as reported by Our World in Data \citep{owid}.  Complete data was available for 26 countries which brings the total number of groups to $K=31$.  The $CC_{k}$ and $T_{k}$ numbers were selected as reported on May 1, 2020 (or the earliest date during the following week for which data was available).  Numbers for $D_{k}$  for $k=6,\ldots, K$, were obtained from 14 days afterwards, to allow for the known delay between the onset of symptoms and death. \textcolor{black}{ (Some early literature (e.g., \citet{wu2020estimating}; \citet{linton2020incubation}) suggests that the median time from symptom onset to death may be longer than 14 days.)}

    Note that our $T_{k}$ numbers are not ideal since some countries report the number of people tested, while others report the total number of tests (which will be higher if a single person is tested several times).  Also note that, as stated in Section \ref{sec:notation_sub}, the $K$ different groups should, in principle, be entirely independent samples.  This is clearly not the case with the European data (case in point: there are three different groups from within Switzerland; $k=2$, $k=5$, and $k=30$).

We included several covariates about each country's population to explain variation in IR and IFR.  Specifically, for the IR, we consider: (1) the number of days since the country reported 10 or more confirmed infections (``Days since outbreak'') (as reported by \citet{oxford}); (2) the number of days between a country's first reported infection and the imposition of social distancing measures (``Days until lockdown'') (calculated based on when the Government Response Stringency Index (GRSI) reached 20 or higher as reported in \citet{owid}); and (3) the population density (``Pop. density'') (as reported by \citet{owid} and other publicly available sources\footnote{For Geneva (\url{https://www.bfs.admin.ch/bfs/en/home/statistics/regional-statistics/regional-portraits-key-figures/cantons/geneva.html}); for Gangelt (\url{https://en.wikipedia.org/wiki/Gangelt}); for Split-Dalmatia (\url{https://en.wikipedia.org/wiki/Split-Dalmatia_County}); for Zurich  (\url{https://www.bfs.admin.ch/bfs/en/home/statistics/regional-statistics/regional-portraits-key-figures/cantons/zurich.html}).}).  For the IFR, we consider: (1) the share of the population that is 70 years and older (``Prop. above 70 y.o.'') (as reported in \cite{ioannidis2020infection} or \citet{owid}); and (2) the number of hospital beds per 1,000 people (``Hosp. beds per 1,000'')\footnote{obtained from \citet{owid}  or from \url{www.bfs.admin.ch} for Geneva and Zurich cantons.}.  Tables \ref{tab:eur} and \ref{tab:eur2} in the Appendix list all the data used in the analysis.  

%

% population density (as reported by \citet{owid} and other publicly available sources\footnote{For Geneva (https://www.bfs.admin.ch/bfs/en/home/statistics/regional-statistics/regional-portraits-key-figures/cantons/geneva.html); for Gangelt (https://en.wikipedia.org/wiki/Gangelt); for Split-Dalmatia (\url{https://en.wikipedia.org/wiki/Split-Dalmatia_County}); for Zurich  (https://www.bfs.admin.ch/bfs/en/home/statistics/regional-statistics/regional-portraits-key-figures/cantons/zurich.html).})
%& &   & &  &  &  & prop. above & hosp. beds &  days since & days till & pop.  \\ 
 %&&& 70 y.o. & per 1,000 &  10 infections  & lockdown & density\\ 

\subsection{Using only seroprevalence studies}

\begin{figure}[h]
    \centering
    \includegraphics[width=13cm]{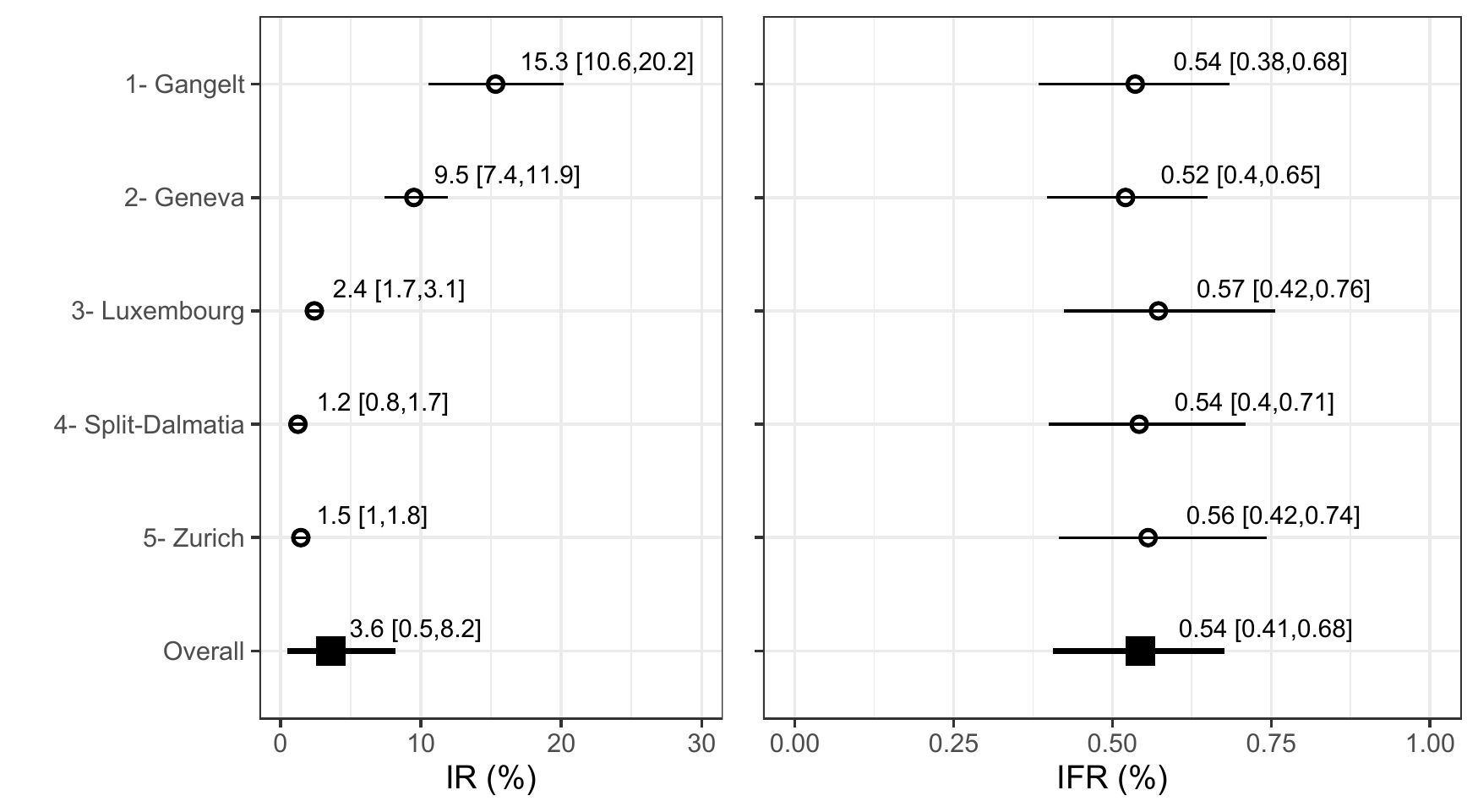}
    \caption{Posterior \textcolor{black}{median estimates for the $IR_{k}$ and $IFR_{k}$ variables (for $k=1,\ldots,5$) with 95\% HPD CIs.  Also plotted, under the label ``Overall'', is the posterior median estimate and 95\% HDP CI of $\operatorname{g}^{-1}(\beta)$ and $\operatorname{g}^{-1}(\theta)$. These results correspond to the large-$P$ model with $\eta=0.1$ which pools information from five seroprevalence studies ($K=5$ and $k^{'}=5$). }
   % R-code to reproduce: \protect\url{https://tinyurl.com/y64wofs3}
    }
    \label{fig:sero}
\end{figure}

Using only the seroprevalence studies (i.e., only the first $k^{'}=5$ studies listed in Table \ref{tab:eur} in the Appendix), we fit the model as described in Section \ref{sec:largeP} (with $\eta=0.1$) without any adjustment for covariates.  (With only $K=5$ groups there are few degrees of freedom available for including group-level covariates).  The model was fit using Stan \citep{carpenter2017stan}, with 4 independent chains, each with 10,000 draws (10\% burn-in, thinning of 50).  Figure \ref{fig:sero} plots the posterior \textcolor{black}{medians} obtained for the $IR_{k}$ and $IFR_{k}$ parameters (for $k=1,\ldots,5$) with 95\% HPD CIs.  We also plot, in black, the posterior median of $\operatorname{g}^{-1}(\beta)$ and $\operatorname{g}^{-1}(\theta)$ (``Overall'').  Our estimate for the overall IFR is $\operatorname{g}^{-1}(\theta)= 0.54\%$,  95\% C.I.    =  [0.41\%,    0.68\%].    \textcolor{black}{We note that the five $IFR_{k}$ estimates obtained are very homogeneous.  This is no doubt partly due to the punitive nature of our prior on $\tau$ (i.e.,  due to setting $\eta=0.1$).  Indeed, when the model is fit with $\eta=1$, the $IFR_{k}$ estimates are more heterogeneous (the posterior median estimates obtained with $\eta=1$ are: $IFR_{k} = 0.50, 0.48, 0.65, 0.53, 0.60$, for $k=1,\ldots,5$, respectively, and $\operatorname{g}^{-1}(\theta)= 0.54\%$,  95\% C.I.    =  [0.37\%,    0.80\%]).}

%This is somewhat similar to what was estimated by \cite{verity2020estimates}, early on during the pandemic: 0.66\% (with 95\% CI = [0.39\%, 1.33\%]).

\subsection{Using all the data}

We fit the model as described in Section \ref{sec:largeP} to all the data (listed in Tables \ref{tab:eur} and \ref{tab:eur2} in the Appendix) with $k^{'}=5$, $K=31$, $h = 3$, and $q = 2$.  Covariates were defined as the centered and scaled logarithm of each metric as follows: $X_{[1]} = \operatorname{center-scale}(\log(\textrm{``Days since outbreak''}))$;   $X_{[2]} = \operatorname{center-scale}(\log(\textrm{``Days until lockdown'' + 1}))$; $X_{[3]} = \operatorname{center-scale}(\log(\textrm{``Population density''}))$; $Z_{[1]} = \operatorname{center-scale}(\log(\textrm{``Prop. above 70 y.o.''}))$; and $Z_{[2]} = \operatorname{center-scale}(\log(\textrm{``Hosp. beds per 1,000''}))$.  Standard normal priors ($\mathcal{N}(0, 1)$) were used for each of $\beta_{1}$, $\beta_{2}$, $\beta_{3}$, $\theta_{1}$, and $\theta_{2}$.  All other priors were defined as in Section \ref{sec:smallP} with $\eta=0.1$ and $\lambda=0.05$.  The model was fit using Stan \citep{carpenter2017stan}, with 4 independent chains, each with 10,000 draws (10\% burn-in, thinning of 50).

\begin{figure}[h]
    \centering
    \includegraphics[width=15cm]{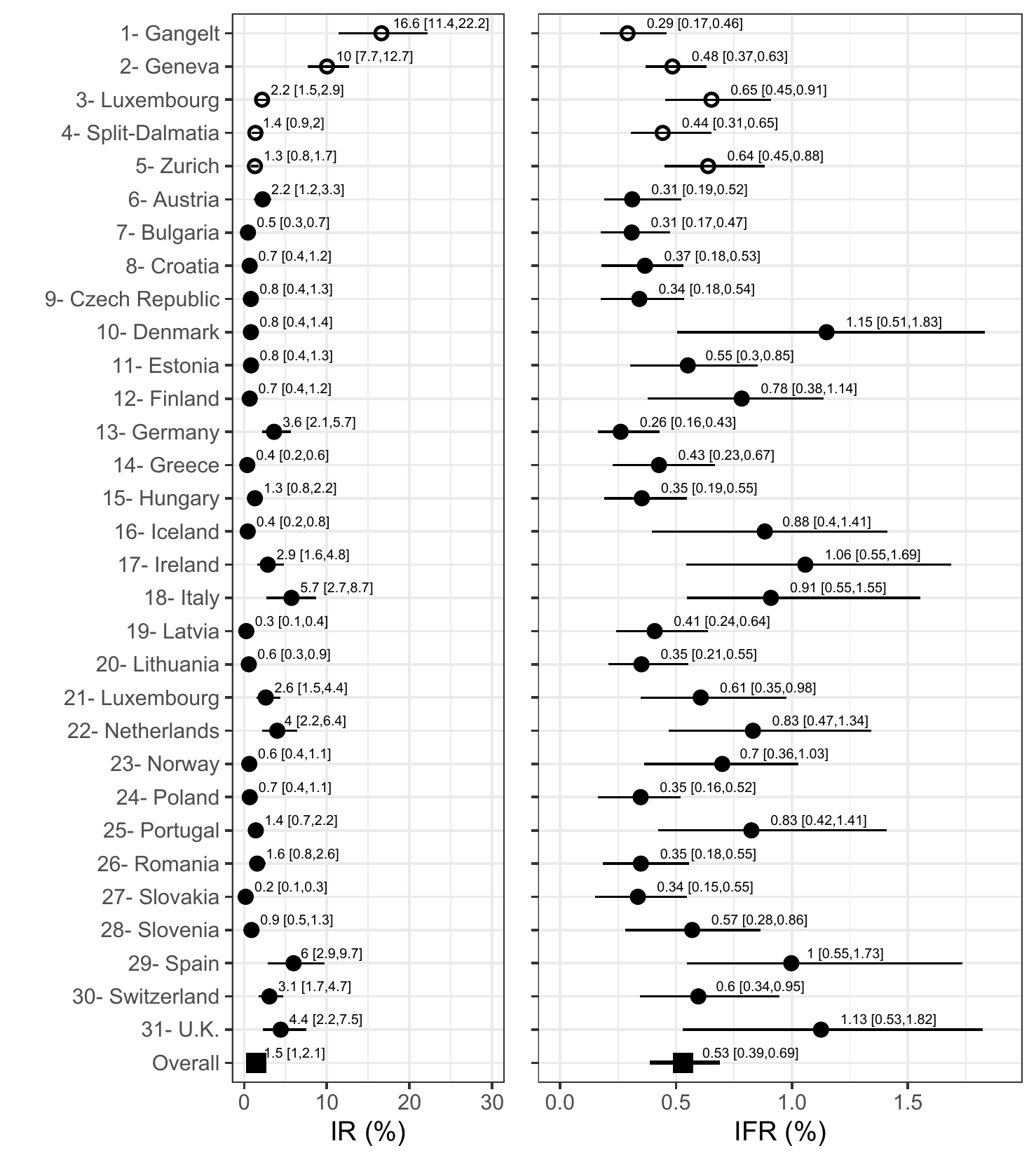}
    \caption{Posterior \textcolor{black}{median estimates for the $IR_{k}$ and $IFR_{k}$ variables (for $k=1,\ldots,31$) with 95\% HPD CIs.  Also plotted, under the label ``Overall'', is the posterior median estimate and 95\% HDP CI of $\operatorname{g}^{-1}(\beta)$ and $\operatorname{g}^{-1}(\theta)$. These results correspond to the large-$P$ model with $\eta=0.1$ which pools information from five seroprevalence studies and data from nationally reported statistics for 26 European countries ($K=31$ and $k^{'}=5$). }  
   %R-code to reproduce: \protect\url{https://tinyurl.com/y64wofs3}
   }
    \label{fig:all}
\end{figure}

\textcolor{black}{Figure \ref{fig:all}  plots the estimates (posterior medians) obtained for the $IR_{k}$ and $IFR_{k}$ variables (for $k=1,\ldots,31$) with 95\% HPD CIs.  We also plot the posterior medians of $\operatorname{g}^{-1}(\beta)$ and $\operatorname{g}^{-1}(\theta)$ (``Overall'').  Our estimate for the overall IFR is  $\operatorname{g}^{-1}(\theta)= 0.53\%$,  95\% C.I.    =  [0.39\%,  0.69\%].}

In the Appendix, Table \ref{tab:coef} lists posterior medians with HPD 95\% CIs for the main parameters of interest.  The positive values for $\beta_{1}$ (0.21, 95\% CI =  [-0.09, 0.54]) and $\beta_{2}$ (0.45,  95\% CI = [0.11, 0.77]) suggest that the IR increases with increasing time since the initial disease outbreak, and with increasing time between the first reported infection and the imposition of social distancing measures.  The positive value for $\beta_{3}$ (0.77, 95\% CI =  [0.50, 1.03]) suggests that a higher population density is associated with a higher IR.  The negative value for $\theta_{2}$ (-0.43, 95\% CI = [-0.63, -0.25]) suggests that countries with fewer hospital beds have higher IFRs.  

\textcolor{black}{We were quite surprised to see that our estimate of $\theta_{1}$ (0.00, 95\% CI = [-0.16, 0.17]) was not decidedly larger and positive given that age is known to be an important predictor of COVID-19 complications and death \citep{o2020age}.   There are several reasons why we might have obtained this result.  First, statistical power may have been compromised by insufficient heterogeneity in the age-structure across different countries, as captured by the proportion aged over 70 metric.  Furthermore, as with a standard multivariable regression of observational data, the estimate of $\theta_{1}$ may suffer from bias due to unobserved confounding and/or multicollinearity.}

\textcolor{black}{Note that the overall IFR estimate and its uncertainty are very similar whether or not the data from nationally reported statistics are included in the analysis (only sero-studies:  $\operatorname{g}^{-1}(\theta)= 0.54\%$,  95\% C.I.    =  [0.41\%,    0.68\%] \textit{vs.} all data:  $\operatorname{g}^{-1}(\theta)= 0.53\%$,  95\% C.I.    =  [0.39\%,  0.69\%]).  One might therefore question what added value the expanded analysis provides.  We have two comments on this point.}

\textcolor{black}{First, since we did not incorporate any substantial prior information about the magnitude of preferential testing into the model (i.e., we selected very ``weakly informative'' priors for the $\phi$ parameters), we should not expect the point estimate of $\theta$ to differ substantially between the two analyses.  In fact, if the difference between the two point estimates was substantial, we might reasonably question whether the priors were appropriately specified.  Since our only unbiased information about the magnitude of the IFR comes from the five seroprevalence studies, a large discrepancy between the two estimates might indicate that our supposedly ``weakly informative'' priors are not as weak as intended.}

\textcolor{black}{Second, even if the estimate for the overall IFR is left unchanged, incorporating the additional data into an expanded analysis is still worthwhile.  Simulation study results (see Section \ref{sec:simresults}) show that a considerable sharpening of information is possible in many scenarios.  Without actually implementing the expanded analysis, it would be impossible to know whether or not such a sharpening would occur in the present context.  Moreover, unlike the analysis which uses only seroprevalence study data, the expanded analysis allows one to obtain valuable country-specific IFR and IR estimates, as well as obtain estimates for the association between the IFR/IR and a number of different explanatory factors.}

%Note that we  repeated our entire analysis setting the prior for $\gamma$ with both $\lambda=0.25$ and $\lambda=0.75$ and obtained estimates for the overall IFR of 0.50\%  (95\% C.I. =  [0.37\%, 0.67\%]) and of 0.44\% (95\% C.I. =  [0.30\%, 0.59\%]), respectively.

%\citet{okell2020have}: ``disparate regions have experienced a similar mortality per infection.'' ``Lithuania is among Europe's leaders in terms of testing''

The model can no doubt be improved by using appropriately specified informed priors for the $\phi$ parameters based on what is known about COVID-19 testing in different countries.  For example, in related work, \citet{grewelle2020estimating} assume that testing capacity is directly proportional to the case load in each country (where testing capacity is estimated by tests performed per positive case\footnote{\citet{grewelle2020estimating} are thereby able to infer the ``global IFR'' using simple weighted linear regression (i.e.,  regressing $\log(D_k/CC_k) \sim \log(IFR_k) + \beta_{1}(CC_k/T_k)$,  for $k$ = $1,\ldots,K$, where $\beta_{1}$ is an unknown nuisance parameter).}). As another example, the ``H2 index'' \citep{oxford},  which purportedly reflects official government policy on who has access to testing within a given country, could also be used to define informed priors for the $\phi$ parameters in a more sophisticated version of our model.

\begin{figure}[h]
    \centering
    \includegraphics[width=15cm]{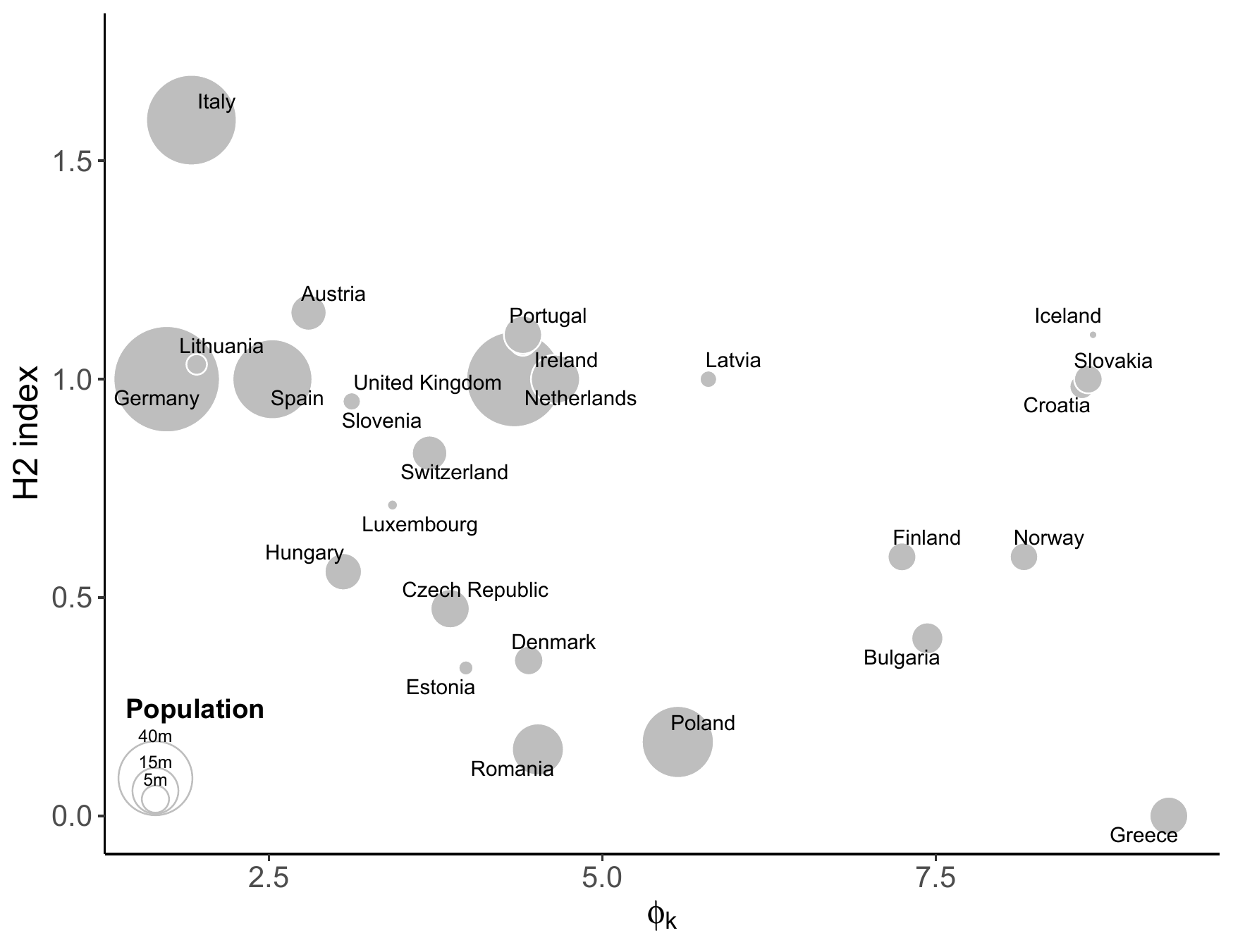}
    \caption{Scatter-plot shows of the average H2 index for each country (for the period between February 1st, 2020 and April 1st, 2020) vs. the posterior median $\phi_{k}$ value.  Circle size corresponds to population ($P_{k}$).  These results correspond to the large-$P$ model with $\eta=0.1$ which pools information from five seroprevalence studies and data from nationally reported statistics for 26 European countries. 
    %R-code to reproduce: \protect\url{https://tinyurl.com/y64wofs3}
    }
    \label{fig:H2}
\end{figure}

We were curious as to whether the model estimates (posterior medians) we obtained for $\phi_{k}$ (for $k$ = $6,\ldots,31$) might be predictive of the H2 index.  Using the data made available by \citet{oxford}, we calculated the average H2 index for each country in our analysis, for the period between February 1st, 2020 and April 1st, 2020.  Roughly speaking, a high H2 value indicates broad access to testing (i.e., available to the general public) whereas a low H2 value reflects a testing policy that restricts testing to only those who have symptoms or meet specific criteria.  Thus, countries with high H2 values should, in theory, have small values of $\phi_{k}$ and vice-versa. That prediction is generally supported by the results seen in Figure \ref{fig:H2}, although Iceland, Slovakia and Croatia appear to be exceptions.

\section{Discussion}
\subsection{Model limitations}

Estimation of the IFR is very challenging due to the fact that it is a ratio of numbers where both the numerator and denominator are subject to a wide range of biases.  Our proposed model seeks to address only one particular type of bias pertaining to the denominator: the bias in the number of cases due to preferential testing.  With this in mind, we wish to call attention to several other important sources of bias.  

%First, one must acknowledge that the definition of ``death from disease'' is not uniform across countries/jurisdictions. 

%, compiled from death certificates, may not list SARS-CoV-2 as a contributing factor and certain jurisdictions may not have adopted the International Form of Medical Certificate of Cause of Death or have adopted the WHO guidelines on registering COVID-19-related deaths \citep{who1}. As such, reported statistics on the number of deaths

Cause of death information may be very inaccurate.  To overcome this issue, many suggest looking to ``excess deaths,'' by comparing aggregate data for all-cause deaths from the time during the pandemic to the years prior \citep{leon2020covid}.  Using this approach and a simple Bayesian binomial model, \citet{rinaldi2020empirical} are able to obtain IFR estimates without relying on official (possibly inaccurate) data for the number of COVID-19 deaths.

Some people who are currently sick will eventually die of the disease, but have not died yet.  Due to the delay between disease onset and death, the number of confirmed and reported COVID-19 deaths at a certain point in time will not reflect the total number of deaths that will occur among those already infected (right-censoring).  This will result in the number of recorded deaths underestimating the true risk of death.  The denominator of the IFR must be the number of cases \emph{with known outcomes}.

%Using time-series survival data or a defensible prior on the time from infection to death, the Bayesian model could be expanded to account for this additional source of bias/uncertainty.

%numerator and 

%Other limitations of the model include the failure to account for potential correlation between population-specific IFR and IR and for potential correlation between degree of ``preferential testing'' and IFR. 

\textcolor{black}{The model assumes that no individuals are tested more than once.  This is an important practical limitation.}  In addition, the model, as currently proposed, fails to account for the (unknown) number of false positive and false negative tests.  When both the test specificity and the infection rate is low, false positives can substantially inflate the estimated infection rate and as a consequence, the IFR could be biased downwards.  In principle, the model could accommodate for this by specifying priors for test sensitivity and specificity; see \citet{Kummerer2020, gelmancarp} and \citet{neil2020}.

%\citet{srinivasan2012propagation, burstyn2020towards, gelmancarp}.   are two recent examples of Bayesian models used to estimate the COVID-19 IFR from serological antibody survey data while accounting for the uncertainty related to testing sensitivity and specificity.

%In fact, Bayesian inference is known to be an excellent tool for adjusting for unknown testing uncertainty 

Finally, because the model uses data that are aggregated at the group level, estimates are potentially subject to ecological bias \citep{pearce2000ecological}.  While including group-level covariates may help reduce variability in the estimates, adjustment using group-level covariates can also lead to misleading results \citep{li2020closer}. %berlin2002individual  %While ecological analyses can be useful for developing hypotheses and may be needed to make rapid use of publicly available surveillance data for inference during an epidemic, they cannot be used to make reliable inferences at the participant level. Sharing de-identified participant-level data as rapidly and widely as possible, in keeping with ethical and legal standards, is central to epidemic response \citep{datasharing}. 

\subsection{Concluding remarks}

\textcolor{black}{Obtaining representative data remains challenging and costly.  Efforts to better understand the distribution of SARS-CoV-2 infection (and its lethality) at the population level have unfortunately been met by recruiting challenges \citep{gudbjartsson2020spread, bendavid2020covid}, leading to an over-representation of people who are concerned about their exposure and/or an under-representation of individuals who are self-quarantining, isolating, or hospitalized because of the virus.   In the absence of large-scale unbiased data, researchers must work with whatever data is available.  Our model suggests a coherent and feasible way to do just this.}

We demonstrated our proposed model with an application to European COVID-19 data, in which we relied on data from seroprevalence studies that self-reported as representative.  When combined with data from nationally reported statistics, this data enabled us to obtain appropriate estimates for not only the overall IFR but also for country-specific IFRs and IRs, as well as for the association between these and various explanatory factors.  

\textcolor{black}{We note that our estimate for the overall IFR in Europe (0.53\%,  95\% C.I.    =  [0.39\%,  0.69\%]) is somewhat lower than an estimate obtained by the meta-analysis of \citet{meyerowitz2020systematic} from European seroprevalence studies (IFR = 0.77\%, 95\% C.I. =  [0.55\%,  0.99\%]) and reiterate that the primary intention of our analysis was to demonstrate the feasibility of the proposed model.  That being said, we selected the five  seroprevalence studies based on the review of \cite{ioannidis2020infection} and suspect that the differing inclusion/exclusion criteria between \cite{ioannidis2020infection} and \citet{meyerowitz2020systematic} are the main reason why our estimates are not more similar.  A recent large-scale analysis for Spain, based on a nationwide population-based seroprevalence study  \citep{pollan2020prevalence} concludes that,  for the Spanish population, ``overall infection fatality risk was 0.8\%  [...] for confirmed COVID-19 deaths and 1.1\% [...] for excess deaths [...]''  \citep{pastor2020infection}.  This is in reasonable agreement  with our estimate for Spain of 1.00\% (95\% credible interval of [0.55\%, 1.73\%]).}  

\textcolor{black}{In a similar study, \citet{o2020age} conduct a Bayesian analysis using data from 22 national-level seroprevalence surveys and sex and age-specific COVID-19-associated death data from 45 countries.  Their data is much more granular and their model specifies many more detailed assumptions.  For example,  \citet{o2020age} assume ``a gamma-distributed delay between onset [of infection] and death'' and assume different risks of infection for ``individuals aged 65 years and older, relative to those under 65'' since ``older individuals have fewer social contacts and are more likely to be isolated through shielding programmes''.    \citet{o2020age} conclude that ``population-weighted IFR estimates by the ensemble model are highest for countries with older populations such as [...]  Italy (0.94\%; 95\% credible interval, 0.80--1.08\%).''  While our model did not identify a significant association between IFR and countries with older populations (much to our surprise), many of our country-specific IFR estimates are quite similar to those reported in  \citet{o2020age}.   For example, we obtained a IFR for Italy of 0.91\% (95\% credible interval of [0.55\%, 1.55\%]).}

In the Introduction, we identified two important questions. First (Q1), is it possible to reliably estimate the IFR without any information about the degree to which the data are biased by preferential testing?   And second (Q2), when representative samples are available, can samples with an unknown degree of preferential testing contribute valuable information?  The proposed Bayesian model suggests that reliable estimation of the IFR at the group level is indeed possible, to a certain extent, when existing data do not arise from a random sample from the target population.  Importantly, the key to (partial) identifiability is sufficient heterogeneity in the degree of preferential testing across groups and sufficient homogeneity in the group-specific IFR.   We also saw that combining both types of data (biased and unbiased data) can be superior to ignoring data that may be skewed by preferential testing.  When fit with an appropriate model, biased data can supplement available representative data in order to refine one's inference \textcolor{black}{and/or} shed light on the impact of explanatory factors.  

In a typical situation of drawing inference from a single biased sample, obtaining appropriate estimates is challenging, if not impossible, without some sort of external validation data.  Intuition suggests that one might only be able to do a sensitivity analysis with respect to the impact of bias.  Indeed, applying prior distributions for the degree of preferential testing and proceeding with Bayesian inference could be regarded as a probabilistic form of sensitivity analysis  \citep{greenland2005multiple}.  What is perhaps less intuitive, and what we demonstrated with the proposed model, is that, if one has multiple samples of biased data, each subject to a different degree of bias, the ``heterogeneity of bias'' can help inform what overall adjustment is required for appropriate inference.   

The aggregation of data from both biased and unbiased samples is a problem that applies to many types of evidence synthesis (and is often overlooked) \citep{de2015four, birrell2018evidence}. In that sense, the solutions we put forward may be more broadly applicable.  Future work should investigate whether the ``heterogeneity of bias'' principle (see Section \ref{sec:unident}) can be used to derive appropriate estimates in a meta-analysis where individual studies are subject to varying degrees of bias due to unobserved confounding or measurement error \textcolor{black}{ (e.g., \citet{campbell2020measurement}).}
\bibliographystyle{apalike}

\bibliography{mix_refs}

\pagebreak

\section{Appendix}
\subsection{Details for the non-central hyper-geometric distribution}
\label{sec:noncentralhyper}

{\color{black}
In Section \ref{sec:notation_sub}, we consider the distribution of $CC|C$ as following Wallenius' {non-central} hyper-geometric (NCHG) distribution such that:
   \begin{align}
   CC |  C &\sim {NCHyperGeo}(C, P-C, T, \phi),
   \label{eq:CC_NCHG}
\end{align}
    \noindent where the degree of preferential testing corresponds to the $\phi$ non-centrality parameter.  The probability mass function of this distribution \citep{lyons1980m29, fog2008sampling} can be written out as:
%
%$m1 = C$
%$m2=P - C$
%
\begin{equation}
f(x) = {\binom {C}{x}}{\binom {P - C}{T-x}}\int _{0}^{1}(1-t^{\phi_{k} /B})^{x}(1-t^{1/B})^{T-x}\operatorname {d} t
\end{equation}
\noindent where $ B=\omega (C-x)+(P - C-(T-x))$.  Note that when the non-centrality parameter, $\phi$, equals 1, the distribution is equivalent to the standard central hypergeometric distribution. 

Wallenius' noncentral hypergeometric is often confused with Fisher's noncentral hypergeometric distribution.   Wallenius' noncentral hypergeometric distribution describes the situation where a predetermined number of items are seleted one by one, whereas  Fisher's noncentral hypergeometric distribution describes a situation where the total number of items drawn is only known after the experiment. }

\subsection{Issues of (un)identifiability}
\label{sec:unident}

Table \ref{tab:ex1} provides a small artificial dataset to help illustrate the type of data being described and the impact of different degrees of preferential testing.  In this dataset, we have $K=12$ groups and the (unknown) infection rate varies substantially  from 13\% to 53\%.  The unknown infection fatality rate only varies slightly, from 0.017\% to 0.022\%.   Values for $\phi_{k}$ in this dataset are evenly distributed between 1 and $\gamma+1$, for four different values of $\gamma = $ 0, 4, 11, and 22.  When $\gamma=0$, the number of true cases (i.e. actual infections) is approximately 14 times higher than the number of confirmed cases.  In contrast, when $\gamma=22$, the number of true cases is only about 5 times higher than the number of confirmed cases. 

%\vspace{1cm}

\begin{table}[ht]
\centering
\begin{footnotesize}
\begin{tabular}{r|rrrrrrr|rrrrrr}
   &  \multicolumn{3}{l} {\textbf{Observed}}  &$\gamma=0$ &$4$ &$11$ &$22$&\multicolumn{3}{l} {\textbf{Unobserved}}   &$\gamma=4$ &$11$  &$22$ \\ 
 $k$ & $P_{k}$ & $T_{k}$ & $D_{k}$ & $CC_{k}$ & $CC_{k}$ & $CC_{k}$ &  $CC_{k}$ &  $C_{k}$  &  $IR_{k}$ &  $IFR_{k}$  &  $\phi_{k}$ & $\phi_{k}$ & $\phi_{k}$ \\ 
  \hline
1 & 3061 & 190 &   11 & 24 & 21 & 32 & 27 & 430 & 0.140 & 0.018 & 1   & 1 & 1  \\ 
  2 & 482 & 43 &    2 & 15 & 11 & 12 & 24 & 99 & 0.206 & 0.020 & 1.36 & 2 & 3  \\ 
  3 & 1882 & 101 &   20 & 32 & 40 & 55 & 74 & 570 & 0.303 & 0.022 & 1.73 & 3 & 5  \\ 
  4 & 1016 & 67 &    2 & 14 & 24 & 33 & 38 & 193 & 0.190 & 0.017 & 2.09 & 4 & 7  \\ 
  5 & 1269 & 109 &    4 & 13 & 34 & 54 & 67 & 201 & 0.159 & 0.021 & 2.45 & 5 & 9  \\ 
  6 & 3670 & 276 &    9 & 53 & 70 & 140 & 162 & 484 & 0.132 & 0.021 & 2.82 & 6 & 11  \\ 
  7 & 2409 & 139 &    7 & 17 & 34 & 70 & 94 & 329 & 0.137 & 0.019 & 3.18 & 7 & 13  \\ 
  8 & 1074 & 81 &   13 & 42 & 65 & 68 & 77 & 565 & 0.526 & 0.019 & 3.55 & 8 & 15  \\ 
  9 & 3868 & 289 &   16 & 60 & 142 & 205 & 247 & 821 & 0.212 & 0.019 & 3.91 & 9 & 17  \\ 
  10 & 151 & 13 &    2 & 1 & 5 & 11 & 8 & 24 & 0.160 & 0.019 & 4.27 & 10 & 19  \\ 
  11 & 430 & 25 &    1 & 6 & 9 & 16 & 18 & 70 & 0.164 & 0.019 & 4.64 & 11 & 21  \\ 
  12 & 429 & 40 &    2 & 11 & 23 & 31 & 33 & 105 & 0.245 & 0.019 & 5   & 12 & 23  \\ 
  \hline
\end{tabular}
\end{footnotesize}
\caption{Illustrative Example Data, with varying degrees of preferential sampling, $\gamma=0$, $\gamma=4$, $\gamma=11$, and $\gamma=22$.  
%R-code to reproduce: \protect\url{https://tinyurl.com/y7gmnpob}
}
\label{tab:ex1}
\end{table}

 Here we present an asymptotic argument which lays bare the flow of information.  Consider a situation in which an infinite amount data are available.  In so-called ``asymptotia,'' we have that populations are approaching infinite size (i.e., for $k$ = 1,$\ldots, K$, we have $P_{k} \rightarrow \infty$), and that the number of tests also approaches infinity (i.e., for $k$ = 1,$\ldots, K$, we have $T_{k} \rightarrow \infty$).  Recall that a hyper-geometric distribution is asymptotically equivalent to a binomial distribution.  As such, we consider the following approximation (as in Section \ref{sec:largeP}): 
 
 $D_{k} \sim Binom(P_{k}, a_{k}) \quad $;  $\quad$ and   $ \quad CC_{k} \sim Binom(T_{k}, b_{k})$, 
 
 \noindent where $a_{k} =  IFR_{k} \times IR_{k}$ and $b_{k} = 1-(1-IR_{k})^{\phi_{k}}$.  
 
 %Both $a_{k}$ and $b_{k}$ are estimated with near-infinite precision for any given value of $\phi_{k}$.  
 
 %(see expressions (\ref{eq:CClarge})-(\ref{eq:D_uncond}) and explanation in Section \ref{sec:largeP})
 
% Note that $a_k$ simply follows from the conditional binomial distribution.  However, this particular {parameterization} of $b_k$ does not emerge from the limit of the non-central hyper-geometric distribution.
% As we discuss in Section \ref{sec:largeP}, we have simply chosen a convenient parameterization for $b_k$ with the important connotation that $\phi_{k}=1$ corresponds to testing at random, but as $\phi_k$ increases, the testing is more preferentially weighted to those truly infected.    
%For example, with an infection rate of $IR_{k} = 0.01$, the binomial sampling probability, $b_k$, is approximately 10 times larger than the infection rate if $\phi_{k}=10$, and about 18 times 
%larger if $\phi_{k}=20$.

% XXX   better interpretation of what  phi means in binomial XXX
 
Presume that the {\em a priori} defensible information about the preferential sampling in the $k$-th group is expressed in the form 
\begin{eqnarray}   \label{eq:phi_bound}
\phi_{k} & \in [\underbar{$\phi$}_{k}, \bar{\phi}_{k}],
\end{eqnarray}
i.e., $\underbar{$\phi$}_{k}$ and $\bar{\phi}_{k}$ are investigator-specified bounds  on the degree of preferential sampling for that jurisdiction.   If one is certain that cases are as likely, or at least as likely, to be tested as non-cases,   $\underbar{$\phi$}_{k}=1$ is appropriate.  If testing is known to be entirely random for the $k$-th group, one would set $\underbar{$\phi$}_{k} = \bar{\phi}_{k}=1$. 

Note that for fixed
$(a_{k},b_{k})$,
$IFR_{k}$ is a function of 
$\phi_{k}$ with the form
 \begin{equation}  \label{eq:IFR(phi)}
 IFR_{k} ( \phi_{k})  = \frac{a_{k}}{ 1-(1-b_{k})^{(1/\phi_{k})}}.
\end{equation}

\noindent Examining (\ref{eq:IFR(phi)}),
knowledge of $(a_{k},b_{k})$, in tandem with (\ref{eq:phi_bound}) restricts the set of possible values for $IFR_{k}$.  In fact it is easy to verify that 
(\ref{eq:IFR(phi)}) is monotone in $\phi_{k}$, hence the restricted set is an interval.    We write this interval as $I_{k}(a_{k},b_{k},\underbar{$\phi$}_{k},\bar{\phi}_{k})$, or simply as $I_{k}$ for brevity.
This is the jurisdiction-specific {\em identification interval} for $IFR_{k}$.
As we approach asymptotia for the $k$-th group, all values inside the interval remain plausible, while all values outside are ruled out; see \citet{manski2003partial}.  This is the essence of the {\em partial identification} inherent to this problem.

Thinking now about the meta-analytic task of combining information, we envision that both $\phi_{k}$ and $IR_{k}$ could exhibit considerable variation  across jurisdictions.    However, the variation in $IFR$ could be small, particularly if sufficient jurisdiction-specific covariates are included (see Section \ref{sec:covariates}).  That is, after adjustment for a jurisdiction's age-distribution, healthcare capacity, and so on, residual variation in $IFR$ could be very modest.  When modeling, we would invoke such an assumption via a prior distribution.  For understanding in asymptotia, however, we simply consider the impact of an {\em a priori} bound on the variability in $IFR$.   Let $\tau$ be the standard deviation of $IFR$ across jurisdictions.  Then we presume 
$\tau$  does not exceed an investigator-specified upper bound of $\bar{\tau}$,  i.e.,  $\tau \leq \bar{\tau}$.

The jurisdiction-specific prior bounds on the extent of preferential sampling, and the prior bound on $IFR$ variation across jurisdictions, along with the limiting signal from the data in the form of $(a,b)$, gives rise to an identification region for 
the average infection fatality rate,
$\overline{IFR} = K^{-1}\sum_{k=1}^{K}IFR_{k}$.
Formally, this interval is defined as
\begin{eqnarray} \label{eq-idef}
I(a, b, \underbar{$\phi$}, \bar{\phi}, \bar{\tau})  =
\left\{ \overline{IFR} : 
\tau \leq \bar{\tau}, 
IFR_{k} \in I_{k}(a_{k},b_{k}, \underbar{$\phi$}_{k},
\bar{\phi}_{k}), \forall k \in \{1, \ldots, K\}
\right\}.
\end{eqnarray}
Again, the interpretation is direct:
in the asymptotic limit, all values of $\overline{IFR}$ inside this interval are compatible with the observed data, and all values outside are not.
The primary question of interest is whether this interval is narrow or wide under realistic scenarios, 
since this governs the extent to which we can learn about $\overline{IFR}$ from the data.

In general, evaluating (\ref{eq-idef}) for given inputs is an exercise in quadratic programming nested within a grid search, hence can be handled with standard numerical optimisation.   
%
%Details of this formulation are given in the Appendix.  
%
However, the special case of $\bar{\tau}=0$ is noteworthy in terms of developing both scientific and mathematical intuition.
Consequently, we explore this case in some depth in what follows.

Scientifically, 
$\bar{\tau}=0$
represents the extreme limit of an {\em a priori} assumption that, 
possibly after covariate adjustment,
$IFR$ is a `biological constant' which does not vary across jurisdictions.
If the prospects for inference are not good when this assumption holds,
they will be even less good under the less strict assumption that the $IFR$ heterogeneity is small, 
but not necessarily zero.
Mathematically, 
the case is much simpler,
with (\ref{eq-idef}) reducing to
\begin{align}
I(a,b,\underbar{$\phi$},\bar{\phi}, 0)  &= \cap_{k}
I_{k} (a_{k}, b_{k}, \underbar{$\phi$}_{k}, \bar{\phi}_{k}).
\end{align} 
As intuition must have it, without heterogeneity, a putative value for the `global' IFR is compatible with the observed data if and only if it is compatible with the data from {\em every} jurisdiction individually.

To illustrate,  consider a scenario with $K=12$ jurisdictions, with a constant infection fatality rate of 2\%, i.e., $IFR_k=0.02$, for $k$ = $1,\ldots,12$.  Say that the infection rates for these jurisdictions lie between 0.132 and 0.526, as per Table 1.  Furthermore, say that the unknown $\phi_k$ values range between $1$ and $23$, as per the rightmost column ($\gamma=22$) of Table 1.

\begin{figure}[h]
    \centering
    \includegraphics[width=12.5cm]{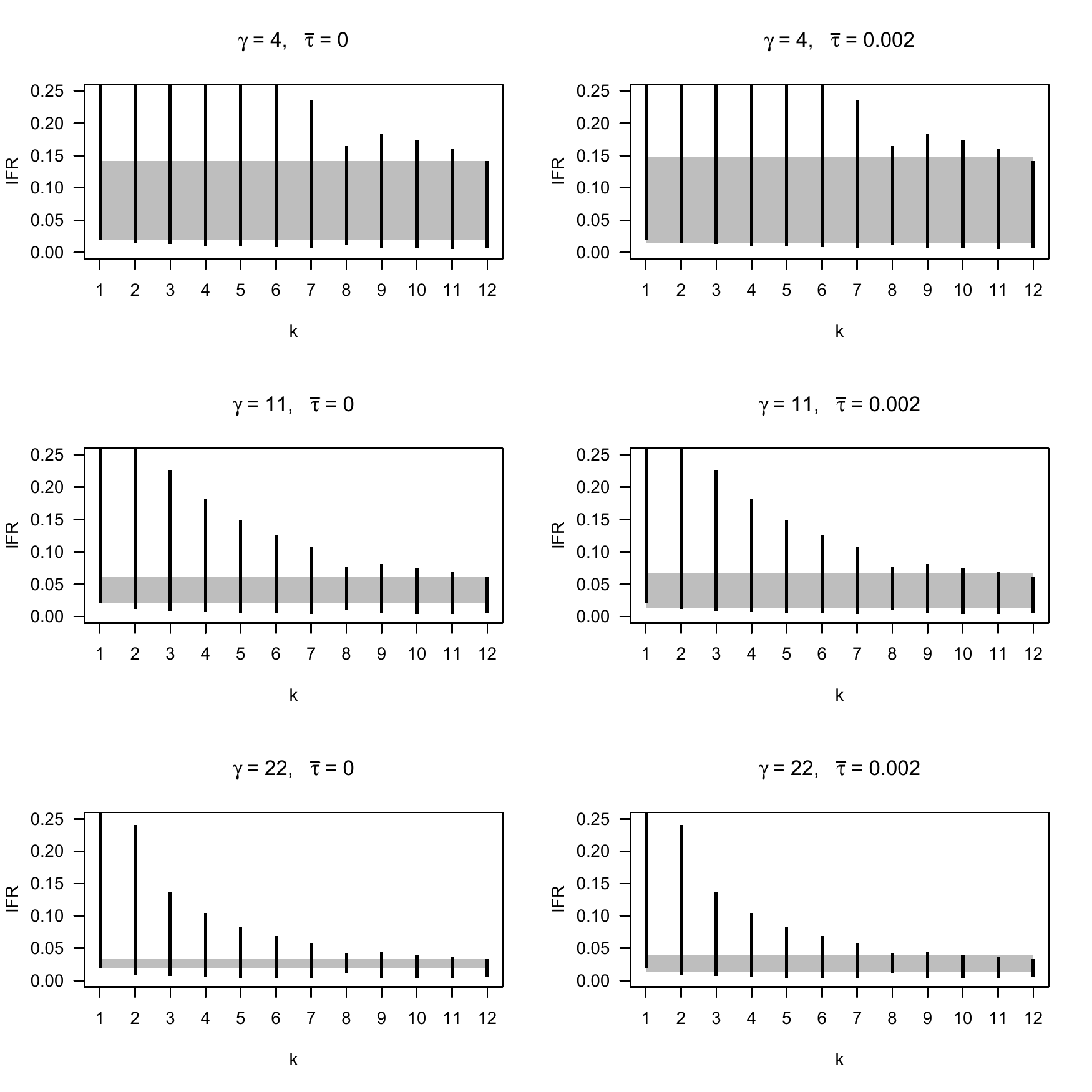}
    \caption{Consider the meta-analytic task of combining information in a scenario with $K=12$ jurisdictions. Here the black lines correspond to jurisdiction-specific identification intervals for the IFR and the grey rectangles corresponds to the global identification interval for IFR.  Left-hand panels correspond to assumption of  $\bar{\tau}=0$ such the global identification interval is simply the intersection of the individual intervals. Right-hand panels correspond to  $\bar{\tau}$ = 0.002. 
    %R-code to reproduce: \protect\url{https://tinyurl.com/yap36pp2}
    }
    \label{fig:greenbox}
\end{figure}

Now say the investigator pre-specifies $(\underbar{$\phi$}_{k}, \bar{\phi}_{k})=(1,40)$ for all $k$. As such, the {\em a priori} bounds are correct, for all jurisdictions.
The resulting jurisdiction-specific identification intervals, $I_k$, are depicted in the bottom left-hand panel of Figure \ref{fig:greenbox}.  (The top and middle left-hand panels correspond to the identical situation but with $\phi_k$ values listed in the $\gamma=4$ and $\gamma=11$ columns of Table 1 respectively.)
Also depicted by the grey rectangle is the global identification interval, i.e., the intersection of the individual intervals.
In the present scenario ($\gamma=22$),  this is indeed narrow, ranging from 0.0200 to 0.0328.  (For the $\gamma=4$, $\gamma=11$ and $\gamma=22$  scenarios, the global identification intervals are [0.0200, 0.1419], [0.0200, 0.0606], and [0.0200, 0.0328], respectively.)  Thus, depending on the range in $\phi_{k}$ values,  i.e., depending on the ``heterogeneity of bias'', it appears that data can contribute substantial information about the (constant) infection fatality rate.

As can be seen immediately from  Figure \ref{fig:greenbox} (left-hand panels), in the present example the binding constraints arise from the first and twelfth  jurisdictions, which happen to have the least and most amounts of preferential testing.  However, this pattern does not hold in general.
One can easily construct pairs of infection rates for which the jurisdiction with more preferential testing has a smaller upper endpoint for $I_k$ and/or a larger lower endpoint.  Thus the values of $\phi_k$ alone do not determine which two jurisdictions will provide the binding information about 
$\overline{IFR}$.

Figure \ref{fig:greenbox} (right-hand panels) shows how the global identification interval is wider when $\overline{\tau}=0.002$.  For reference, for the IFR values listed in Table 1, $\tau=SD(IFR_{1:12})=0.00124$.  For the $\gamma=4$, $\gamma=11$, and $\gamma=22$  scenarios, the global identification intervals outlined by the grey rectangles are [0.0139, 0.1483], [0.0137, 0.0670] and [0.0137, 0.0386], respectively. 

Now consider the evaluation of  (\ref{eq-idef}) for $\overline{\tau}>0$, i.e., where a limited heterogeneity in $IFR$ is permitted.  Recall that quadratic programming constitutes the minimization of a quadratic function subject to linear constraints, and these may be a mix of equality and inequality constraints.   Let $x$ be a candidate value, which we will test for membership in the identification interval.        
To perform this test,
we use a standard quadratic programming package (quadprog)
to minimize the quadratic function $Var(IFR)$, subject to the equality constraint $\overline{IFR}=x$ and the $2K$ inequality constraints which restrict $IFR_{k}$ to the interval $I_k$ for each $k$.
By the definition of (6) then, $x$ belongs in the identification interval if and only if the minimized variance does not exceed $\overline{\tau}^{2}$.

Thus a simple grid search over values of $x$ numerically determines the 
identification interval.
Note that so long as $a$ and $b$ arise from values of $\phi$ within the prescribed bounds, the underlying value of $\overline{IFR}$ must belong to the identification interval.   
Thus two numerical searches can be undertaken.  One starts at the underlying value and tests successively larger $x$ until a failing value is obtained.   The other starts at the underlying value and does the same, but moving downwards.

\subsection{nimble MCMC details}
\label{sec:nimble} 
  
Using nimble, we applied two sampling strategies for the trio ($\phi_{k}$, $\cIRk$, $\cIFRk$) for each $k$ = 1,$\ldots, K$.  In all cases the univariate sampling method was adaptive random-walk Metropolis-Hastings.  For notation, we drop the subscript $k$ and define $\kappa_1 = \cIRk$ and $\kappa_{2} = \cIFRk$.  

First, we included a block sampler on ($\phi$, $\kappa_1$, $\kappa_2$) for each $k$, along with the usual univariate samplers on each element of the trio.  Second, we included samplers in two transformed coordinate spaces.  Define transformed coordinates $(z_1, z_2) = (h_1(\kappa_1, \kappa_2), h_2(\kappa_1, \kappa_2)) = (\exp(\kappa_1) + \exp(\kappa_2), \exp(\kappa_1) - \exp(\kappa_2))$.  (Based on the cloglog link, the quantities $\exp(\kappa_1)$ and $\exp(\kappa_2)$ may be interpreted as continuous time rates.)

Now $z_1$ represents the more strongly identified quantity, so mixing in $z_2$ can be slow.  Hence we wish to improve mixing in the $z_2$ direction.  To do so, a sampler can operate in the $(z_1, z_2)$ coordinates while transforming the prior such that it is equivalent in $(z_1, z_2)$ to what was specified in the original coordinates, $(\kappa_1, \kappa_2)$.  Using $P(\cdot)$ for priors, we have $\log(P(z_1, z_2)) = \log( P(\kappa_1, \kappa_2)) - \log( | J |)$, where $| J |$ is the determinant of the Jacobian of $(z_1, z_2)$ with respect to $(\kappa_1, \kappa_2)$.  In this case, $| J | = 2 \exp( \kappa_1 + \kappa_2 )$.

The other transformed coordinates used were $(z_1, z_2) = (\log(\phi) + \kappa_1, \log( \phi) - \kappa_1)$.  Note that $\log( \phi) + \kappa_1 = \log (-\log ((1-\mbox{IR})^{\phi} ) )$.  Hence $z_1$ represents the more strongly  identified quantity, so we wish to improve mixing by sampling in the $z_2$ direction.  We have the same formulation as above, with $|J| =  2 / \phi$.

\subsection{Seroprevalence study data}
\label{sec:serodetails}

Consider the data for $k=1,\ldots,5$:  

\begin{itemize}

    \item \textbf{$k=1$: Gangelt, Germany}- \citet{streeck2020infection} estimated the infection prevalence from a ``random population sample'' obtained between March 31st, 2020 and April 6th, and provide a 95\% CI for the IR of [12.31\%, 24.40\%].\footnote{\citet{streeck2020infection} reports two different 95\% CIs, obtained with and without applying a ``correction factor'': [15.84\%; 24.40\%] and [12.31\%; 18.96\%], respectively.}  \textcolor{black}{This uncertainty interval is equivalent to a binomial distribution with 27 confirmed cases from 153 tests.  The relevant number of deaths is 8 ( ``until April 20th''), as listed by \citet{streeck2020infection}.}  With a total population of 12,597, this corresponds to a 95\% HPD credible interval for the IFR of [0.15\%,  0.74\%] (see R code below in Section \ref{sec:rcodesero} for this calculation).

    \item \textbf{$k=2$: Geneva, Switzerland}- Based on the data collected by \citet{stringhini2020seroprevalence}, \citet{perez2020serology} estimate a 95\% CI for the IR of [8.15\%, 13.95\%] for a ``representative sample of the general population'' of the canton of Geneva (enrollment between April 6th and May 9th).\footnote{Population number for the canton of Geneva obtained from \citet{perez2020serology}.}  This uncertainty interval is equivalent to a binomial distribution with 48 confirmed cases amongst 442 tests.  The relevant number of deaths is 243 (recorded on April 30th, 2020), as listed by \citet{ioannidis2020infection}.\footnote{\label{note3}\citet{ioannidis2020infection}: ``For the number of COVID-19 deaths, the number of deaths recorded at the time chosen by the authors of each study was selected, whenever the authors used such a death count up to a specific date to make inferences themselves. If the choice of date had not been done by the authors, the number of deaths accumulated until after 1 week of the mid-point of the study period was chosen.''} With a total population of 506,765, this corresponds toa 95\% HPD credible interval for the IFR of [0.32\%, 0.59\%].
   
    \item \textbf{$k=3$: Luxembourg}: \citet{snoeck2020prevalence}  ``recruited a representative sample of the Luxembourgish population'' between April 16th and May 5th, and obtained a 95\% CI of [1.23\%, 2.77\%].\footnote{\citet{snoeck2020prevalence} report two different 95\% CIs, obtained with and without adjustment for age, gender and canton: [1.23\%; 2.67\%] and  [1.34\%; 2.77\%].}  This uncertainty interval corresponds to about 23 confirmed cases from 1,214 tests.  The relevant number of deaths is 92 (recorded on May 2nd, 2020), as listed by \citet{ioannidis2020infection}.  With a total population of 615,729,  this corresponds to a 95\% HPD credible interval for the IFR of  [0.48\%, 1.24\%].
    
    \item \textbf{$k=4$: Split-Dalmatia County, Croatia}: \citet{jerkovic2020sars}   conducted serological testing for antibodies from April 23rd to April 28th, and obtained a 95\% CI for the IR of [0.64\%,	2.05\%] (from ``a representative sample size for the Split-Dalmatia County population,  which could reflect a relatively realistic antibody seroprevalence in the county'').  This uncertainty interval corresponds to about 12 confirmed cases from 938 tests.  The relevant number of deaths is 29 (recorded on May 3rd, 2020), as listed by the Croatian Institute of Public Health (www.koronavirus.hr).  With a total population of 447,723, this corresponds to a 95\% HPD credible interval for the IFR of [0.24\%, 0.99\%].
    
    \item \textbf{$k=5$: Zurich, Switzerland} (May, 2020): \citet{emmenegger2020early} estimate a 95\% CI for the IR of  [0.6\%, 1.8\%] for ``the first half of April 2020'' and note that ``the prevalence reported here is truly representative of the population under study.''  This uncertainty interval corresponds to about 13 confirmed cases from 1,167 tests. The relevant number of deaths is 127 (recorded on May 15th, 2020), as listed by \citet{ioannidis2020infection}.  With a total population of 1,520,968, this corresponds to a 95\% HPD credible interval for the IFR of [0.40\%, 1.32\%].
    
\end{itemize}

\pagebreak 

{\color{black}
\subsection{Example R-code for the Gangelt, Germany seroprevalence study data}
\label{sec:rcodesero}

\begin{footnotesize}
\begin{verbatim}
###########
library(rriskDistributions); library(rjags); library("hBayesDM")
# k = 1: Gangelt, Germany
Dk <- 8;  Pk <- 12597; IR_CI <- c(0.1231, 0.2440)
ab_param <- round(get.beta.par(p = c(0.025,0.975), q = IR_CI, plot = FALSE))

modelIFR <- "model {
cc ~ dbin(ir, tests);
deaths ~ dbin(ifr*ir, pop);
ifr ~ dunif(0,1);
ir ~ dunif(0,1)}"

jags.model <- jags.model(textConnection(modelIFR), 
      data = list(
             cc = ab_param[1],
             tests = ab_param[1] + ab_param[2] + 1,
             pop = Pk,
             deaths = Dk))
sm <- coda.samples(jags.model, "ifr", n.iter = 10000000,  thin = 100)
100*round(HDIofMCMC(unlist(sm[,"ifr"])), 4)
#  0.15 0.74
###########
\end{verbatim}
\end{footnotesize}}

\subsection{Tables}
\label{sec:datatables}

% latex table generated in R 3.5.3 by xtable 1.8-4 package
% Wed Jul  8 14:58:47 2020
\begin{table}[p]
\centering
\begin{footnotesize}
\begin{tabular}{rlrrrrr}
  \hline
$k$ & Location & Date & $T_{k}$ & $CC_{k}$ & $P_{k}$ & $D_{k}$ \\
&  & (MM-DD) &  & &  &  \\
  \hline
1 & Gangelt (Germany)  &  04-02 & 153* & 27* & 12597 & 8 \\ 
  2 & Geneva (Switzerland) &  04-23 & 442* & 48* & 506765 & 243 \\ 
  3 & Luxembourg &  04-26 & 1214* & 23* & 615729 & 92 \\ 
  4 & Split-Dalmatia (Croatia) &  04-25 & 938* & 12* & 447723 & 29 \\ 
  5 & Zurich (Switzerland) &  04-07 & 1167* & 13* & 1520968 & 127 \\ 
  6 & Luxembourg &  05-01 & 44895 & 3784 & 625976 & 103 \\ 
  7 & Austria &  05-01 & 264079 & 15424 & 9006400 & 626 \\ 
  8 & Bulgaria &  05-01 & 46510 & 1506 & 6948450 & 99 \\ 
  9 & Croatia &  05-02 & 37557 & 2085 & 4105268 & 95 \\ 
  10 & Czech Republic &  05-01 & 258368 & 7682 & 10708978 & 293 \\ 
  11 & Denmark &  05-01 & 266124 & 9158 & 5792203 & 538 \\ 
  12 & Estonia &  05-01 & 54439 & 1689 & 1326539 & 62 \\ 
  13 & Finland &  05-01 & 106438 & 4995 & 5540716 & 287 \\ 
  14 & Germany &  05-03 & 2773432 & 162496 & 83783930 & 7914 \\ 
  15 & Greece &  05-01 & 77251 & 2591 & 10423036 & 156 \\ 
  16 & Hungary &  05-01 & 76331 & 2863 & 9660352 & 442 \\ 
  17 & Iceland &  05-01 & 49961 & 1797 & 341250 & 10 \\ 
  18 & Ireland &  05-01 & 177097 & 20612 & 4937796 & 1506 \\ 
  19 & Italy &  05-01 & 2053425 & 205463 & 60461823 & 31368 \\ 
  20 & Latvia &  05-01 & 61120 & 858 & 1886203 & 19 \\ 
  21 & Lithuania &  05-01 & 132768 & 1385 & 2722289 & 54 \\ 
  22 & Netherlands &  05-03 & 238672 & 40236 & 17134870 & 5670 \\ 
  23 & Norway &  05-01 & 156444 & 7710 & 5421243 & 232 \\ 
  24 & Poland &  05-01 & 354628 & 12877 & 37846592 & 883 \\ 
  25 & Portugal &  05-01 & 439890 & 24987 & 10196707 & 1184 \\ 
  26 & Romania &  05-01 & 183688 & 12240 & 19237691 & 1046 \\ 
  27 & Slovakia &  05-01 & 91072 & 1396 & 5459651 & 27 \\ 
  28 & Slovenia &  05-01 & 55020 & 1429 & 2078933 & 103 \\ 
  29 & Spain &  05-07 & 1625211 & 222045 & 46754781 & 27940 \\ 
  30 & Switzerland &  05-01 & 280735 & 29503 & 8654617 & 1588 \\ 
  31 & United Kingdom &  05-01 & 996826 & 171253 & 67886017 & 33614 \\ 
   \hline
\end{tabular}
\end{footnotesize}
\caption{{*}$CC_{k}$ and $T_{k}$ numbers for $k=1,\ldots,5$ were obtained by inverting binomial confidence intervals so as to match the reported 95\% CI for the estimated IRs published in the seroprevalence studies.  We assume that $\phi_{k}=1$, for $k=1,\ldots,5$, (seroprevalence studies); and that $\phi_{k}$ is unknown for $k=5,\ldots,31$ (national official statistics).  The date listed for each group corresponds to the date (or midpoint during the study time period) corresponding to the $CC_{k}$ and $T_{k}$ numbers. 
%R-code to reproduce: \protect\url{https://tinyurl.com/y64wofs3}
}
\label{tab:eur}
\end{table}

% latex table generated in R 3.5.3 by xtable 1.8-4 package
% Mon Jul 20 19:00:39 2020
\begin{table}[p]
\centering
\begin{footnotesize}
\begin{tabular}{rlrrrrr}
  \hline
 $k$ & Location & Prop. above & Hosp. beds &  Days since & Days until & Pop. density  \\ 
 && 70 y.o. (\%) & per 1,000 &  outbreak  & lockdown &  (per $km^{2}$)\\ 
  \hline
1 & Gangelt (Germany)  & 14 & 8.00 & 58 & 32 & 260 \\ 
  2 & Geneva (Switzerland) & 12 & 4.80 & 54 & 6 & 2032 \\ 
  3 & Luxembourg & 10 & 4.51 & 44 & 8 & 231 \\ 
  4 & Split-Dalmatia (Croatia) & 14 & 5.54 & 50 & 14 & 100 \\ 
  5 & Zurich (Switzerland) & 12 & 4.10 & 38 & 6 & 916 \\ 
  6 & Austria & 14 & 7.37 & 61 & 14 & 107 \\ 
  7 & Bulgaria & 13 & 7.45 & 49 & 0 & 65 \\ 
  8 & Croatia & 13 & 5.54 & 57 & 14 & 74 \\ 
  9 & Czech Republic & 12 & 6.63 & 56 & 8 & 137 \\ 
  10 & Denmark & 12 & 2.50 & 57 & 5 & 137 \\ 
  11 & Estonia & 13 & 4.69 & 55 & 13 & 31 \\ 
  12 & Finland & 13 & 3.28 & 56 & 42 & 18 \\ 
  13 & Germany & 16 & 8.00 & 89 & 32 & 237 \\ 
  14 & Greece & 15 & 4.21 & 57 & 7 & 83 \\ 
  15 & Hungary & 12 & 7.02 & 51 & 6 & 108 \\ 
  16 & Iceland & 9 & 2.91 & 58 & 15 & 3 \\ 
  17 & Ireland & 9 & 2.96 & 56 & 11 & 70 \\ 
  18 & Italy & 16 & 3.18 & 69 & 21 & 206 \\ 
  19 & Latvia & 14 & 5.57 & 50 & 10 & 31 \\ 
  20 & Lithuania & 14 & 6.56 & 46 & 13 & 45 \\ 
  21 & Luxembourg & 10 & 4.51 & 49 & 8 & 231 \\ 
  22 & Netherlands & 12 & 3.32 & 62 & 13 & 509 \\ 
  23 & Norway & 11 & 3.60 & 61 & 13 & 14 \\ 
  24 & Poland & 10 & 6.62 & 53 & 6 & 124 \\ 
  25 & Portugal & 15 & 3.39 & 55 & 7 & 112 \\ 
  26 & Romania & 12 & 6.89 & 54 & 10 & 85 \\ 
  27 & Slovakia & 9 & 5.82 & 50 & 2 & 113 \\ 
  28 & Slovenia & 13 & 4.50 & 54 & 6 & 103 \\ 
  29 & Spain & 14 & 2.97 & 70 & 37 & 93 \\ 
  30 & Switzerland & 13 & 4.53 & 62 & 6 & 214 \\ 
  31 & United Kingdom & 13 & 2.54 & 67 & 46 & 273 \\ 
   \hline
\end{tabular}
\end{footnotesize}
\caption{Covariate data for the European dataset includes: the share of the population that is 70 years and older (``Prop. above 70 y.o.''), the number of hospital beds per 1,000 people (``Hosp. beds per 1,000''), the number days since the country reported 10 or more confirmed infections (``Days since outbreak''), the number of days between a country's first reported infection and the imposition of social distancing measures (``Days until lockdown''), and the population density (``Pop. density'').}
\label{tab:eur2}
\end{table}

% latex table generated in R 3.5.3 by xtable 1.8-4 package
% Wed Jul  8 15:37:12 2020
\begin{table}[ht]
\centering
\begin{footnotesize}
\begin{tabular}{l|cl|rl}
  \hline
  & \multicolumn{2}{c}{\textbf{Seroprevalence data}} &  \multicolumn{2}{c}{\textbf{All data }} \\
 & Estimate & 95\% CI   & Estimate & 95\% CI \\ 
  \hline
  $\operatorname{g}^{-1}(\theta)$ & 0.005 & [0.004, 0.007] & 0.005 & [0.004,  0.007] \\
  $\operatorname{g}^{-1}(\beta)$ & 0.036 & [0.005, 0.082] & 0.015 & [0.010, 0.021]\\
$\theta$ &  -5.211 & [-5.438, -4.937] &  -5.235 & [-5.527 , -4.950] \\ 
 $\beta$ &  -3.305 & [-4.335, -2.096] & -4.218 & [-4.572 , -3.841] \\ 
  $\theta_{1}$ (``Prop. above 70 y.o.'')& & &  -0.003 & [-0.158 , 0.165] \\ 
  $\theta_{2}$ (``Hosp. beds per 1,000'') && & -0.428 & [-0.630 , -0.244] \\
  $\beta_{1}$ (``Days since outbreak'')  && & 0.211 & [-0.090 , 0.540] \\
  $\beta_{2}$ (``Days until lockdown'') && &   0.451 & [0.113 , 0.773] \\
  $\beta_{3}$ (``Pop. density'') && &  0.766 & [0.496 , 1.031] \\ 
  $\tau$ &    0.081  & [0.002, 0.200] & 0.195 & [0.039 , 0.350] \\
  $\sigma$ &  1.156 & [0.649, 2.057] & 0.683 & [0.502 , 0.923] \\
  $\gamma$ & &  & 9.183 & [4.759 , 14.204] \\ 
   \hline
\end{tabular}
\end{footnotesize}
\caption{Posterior parameter estimates (posterior medians and 95\% HPD CIs) from large-$P$ model fit to data from only the seroprevalence studies (left) and from the full dataset (right).  The large-$P$ model is fit with priors specified by $\lambda=0.05$ and $\eta=0.1$.}
\label{tab:coef}
\end{table}

\end{document}